\documentclass[journal, twocolumn]{IEEEtran}

\usepackage{amsmath,amssymb,graphicx,epsfig,fancyhdr,amsthm,tabulary,epstopdf,stmaryrd}
\usepackage{hyperref}
\usepackage{cite}
\usepackage{multirow}
\usepackage{mpmulti}
\usepackage{float}
\usepackage{amsfonts}
\usepackage{subfigure}
\usepackage{footnote}
\makesavenoteenv{tabular}
\usepackage{tabulary,bigstrut,array,multirow,booktabs}%
\usepackage{tikz}
\usetikzlibrary{calc}
\usepackage{bm}

\usepackage{multirow, cite}
\usepackage{xcolor,colortbl, soul}
\usepackage{tabularx,booktabs}

\usepackage{color, colortbl}
\definecolor{Gray}{gray}{0.9}
\definecolor{lightcyan}{rgb}{0.88,1,1}
\definecolor{cyan}{rgb}{0,1,1}
\definecolor{electricblue}{rgb}{0.49, 0.98, 1.0}
\definecolor{myblue}{rgb}{0.74, 0.98, 1.0}
\definecolor{lightblue}{rgb}{0.68, 0.85, 0.9}

\theoremstyle{remark}

\theoremstyle{definition}

\begin{document}

\title{Interplay Between NOMA and Other Emerging Technologies: A Survey}

\author{\IEEEauthorblockN{Mojtaba Vaezi,   \IEEEmembership{Senior Member, IEEE}, Gayan Amarasuriya,
		\IEEEmembership{Member, IEEE}, Yuanwei Liu,
		 \IEEEmembership{Member, IEEE}, \\ 
		 Ahmed Arafa,  \IEEEmembership{Member, IEEE}, Fang Fang, \IEEEmembership{Member, IEEE}, and Zhiguo Ding, \IEEEmembership{Senior Member, IEEE}}
	 \\ 	 \IEEEmembership{(Invited Paper)}

}

\maketitle


\begin{abstract}
Non-orthogonal multiple access (NOMA) has been widely recognized as a promising way to scale up the number of users, enhance the spectral efficiency, and improve the user-fairness in wireless networks, by allowing more than one user to share one wireless resource. NOMA can be flexibly combined with many existing wireless technologies and emerging ones including multiple-input multiple-output (MIMO),  massive MIMO, millimeter wave communications, cognitive and cooperative communications, visible light communications, physical layer security, energy harvesting, wireless caching, and so on. Combination of NOMA with these technologies can further increase scalability, spectral efficiency, energy efficiency, and greenness of future communication networks. This paper provides a comprehensive survey of the interplay between NOMA and the above technologies. The emphasis is on how the above techniques can benefit from NOMA and vice versa. Moreover, challenges and future research directions are identified.
\end{abstract}

\section{Introduction}
{\let\thefootnote\relax\footnotetext{

M. Vaezi is with the Department of Electrical  and Computer Engineering, Villanova University, Villanova, PA, USA (e-mail: mvaezi@villanova.edu).

G. Amarasuriya is with the Department of Electrical  and Computer Engineering, Southern Illinois University, Carbondale, IL, USA 62901 (email: gayan.baduge@siu.edu).

Y. Liu is with School of Electronic Engineering and Computer Science, Queen Mary University of London,
London E1 4NS, U.K. (e-mail: yuanwei.liu@qmul.ac.uk).  

A. Arafa is with the Department of Electrical and Computer Engineering, University of North Carolina Charlotte, NC, USA (e-mail: aarafa@uncc.edu).

F. Fang and Z. Ding are with the School of Electrical and Electronic Engineering, the University of Manchester, Manchester M13 9PL, U.K. (email:\{fang.fang,zhiguo.ding\}@manchester.ac.uk).

The work of Z. Ding was supported by the UK EPSRC under grant number EP/P009719/2 and by H2020-MSCA-RISE-2015 under grant number 690750.}

\author{XXX   
\thanks{XXX}} 

Multiple access techniques allow multiple users to share the
same communication resource and are instrumental in making cellular communication work \cite{vaezi2018book}.
The first  to the fourth generation (1G
to 4G)  cellular networks are distinguished by
their multiple access methods, yet these radio access methods  have been developed with one common
concept: to \textit{orthogonalize} the signal of different users by allocating
distinct resources (be it frequency, time, code, or space) to different users. The
 orthogonalization of the wireless resources simplifies the receiver design since it
 prevents inter-user interference. Such access methods are not, however, optimal
 theoretically.
 More importantly, they restrict the numbers of users to the number of orthogonal  resources.

Fifth generation (5G) wireless networks must support a large  number of connections with diverse requirements in terms of  throughput and latency.
The need for massive connectivity in 5G   networks and beyond is mainly pushed by the proliferation  of the Internet of things (IoT) devices, which is   projected to have
 20\% to 30\%    annual growth in the next several years.
In view of such projections, the 5G performance requirements set by the international telecommunication union (ITU)  for IMT-2020 
 requires 100 times more connection density than that in 4G.

In order to fulfill the diverse requirements of
 5G and beyond cellular networks, several new technologies have
 been developed during the past decade. Among them is \textit{non-orthogonal multiple access} (NOMA)
 \cite{vaezi2018book, saito2013non,ding2017application}, which
can help to address the above challenges  more efficiently than the conventional orthogonal multiple access (OMA) schemes.
NOMA can be flexibly  combined with many other existing and emerging
technologies, such as
  multiple-input multiple-output  (MIMO) and massive MIMO,  millimeter wave, cognitive and cooperative communications, physical layer security, visible light
communications, energy harvesting, mobile edge computing, etc. NOMA can be combined with these technologies   to further increase the number of users and enhance the system performance in various senses.

This survey paper reviews the  contributions that combine NOMA with the above-mentioned technologies
with an emphasis on how these technologies interplay and benefit from each other.
To better understand these relations,  we first briefly
explain the salient features  of each technology in the following.

			\begin{table*} [!h] 
					\caption {The organization of the paper}
						\begin{center}{ 
	 \begin{tabular}{ |l|l|c|c|c| }
	\hline
	\rowcolor{Gray}  \textbf{Organization}  & \textbf{Technology combined with NOMA}  & \textbf{Figures}  & \textbf{Tables}  \\
	\hline \hline
	\rowcolor{myblue}
 Section~\ref{sec:massive} & Massive MIMO& Fig.~\ref{fig:MMIMO_NOMA_System_Model} & Table~\ref{table:table_1}\\ \hline
  Section~\ref{sec:mm} & mmWave communications&   & Table~\ref{table:table_2} \\ \hline
	\rowcolor{myblue}
		\rowcolor{myblue}
 Section~\ref{sec:co} & 	Cooperative  communications&  & Table~\ref{table:cop}\\ \hline
  Section~\ref{sec:cog} & 	Cognitive communications& Fig.~\ref{fig:CRNOMA} &  \\ \hline
	\rowcolor{myblue}
	  Section~\ref{sec:pls} &  Physical layer  security & Fig.~\ref{fig:pls}& Table~\ref{table:pls}\\ \hline
  Section~\ref{sec:eh} &  Energy harvesting &Fig.~\ref{fig:noma_swipt} and Fig.~\ref{fig:noma_wpcn}& Table~\ref{table_eh}\\ \hline
	\rowcolor{myblue}
 Section~\ref{sec:vlc} & Visible light communications & Fig.~\ref{fig_noma_vlc}& Table~\ref{table_vlc}\\ \hline
  Section~\ref{sec:mec} & Mobile edge computing  & Fig.~\ref{fig:mec}& Table~\ref{table:mec}\\ \hline
\end{tabular} \label{table:org}
}	\end{center} 

	\end{table*}

\subsection{Non-orthogonal Multiple Access (NOMA)}

By allowing multiple users to simultaneously access the same wireless resources,  NOMA   offers a  promising solution to the need for   massive connectivity in 5G and beyond.
This term was coined by Saito \textit{et al.} \cite{saito2013non}, where the authors showed that simultaneous transmission of users' signal  can improve system throughput
and  user-fairness over a single-input single-output (SISO) channel when using orthogonal frequency-division multiple access (OFDMA).
In  theory, however, since several decades ago,
it has been known that concurrent (non-orthogonal) transmission of users messages is  the optimal transmission strategy.
In fact,  to achieve  the  capacity region of the downlink  transmission in a single-cell wireless networks,  modeled by the \textit{broadcast
	channel} (BC), the users must transmit
at the same time and frequency \cite{Cover,ElGamal2011network,tse2005fundamentals}.\footnote{
	Similarly, to obtain the highest achievable region
	in the multi-cell systems, concurrent non-orthogonal transmission is
	required  \cite{HK, vaezi2016simplified,  shin2017non, tse2005fundamentals,ElGamal2011network},
	and orthogonal transmission is suboptimal.	
}
The capacity region
of this channel is achieved using \textit{superposition coding} at the base station (BS). For decoding, the user with
a stronger channel gain (typically the one closer to the BS) uses \textit{successive interference cancellation} (SIC)
to decode its signal free of interference, while the user with weaker channel gain treats the signals of the stronger users as noise. Despite its well-established theory, NOMA community has been prone to several myths and  misunderstandings \cite{vaezi2018non}.

Today, NOMA is actively being
investigated by academia, standardization  bodies, and  industry \cite{NOMA3GPP}.
This  owes, partly, to  the  advances in processing power which make interference cancellation at  user equipment viable.
It is also pushed by the need for massive connectivity  and better spectral efficiency.
The successful operation of this technique, however, depends on knowledge of the channel
state information (CSI) at the BS and end users.
While recent advances in processor capabilities have made SIC, and consequently  NOMA, feasible,
significant research   challenges remain to be addressed before NOMA can be deployed commercially.

%
%
%

%

\subsection{Other Emerging Technologies}
\label{sec:tech}
\subsubsection{Massive MIMO}

Massive MIMO can drastically increase the spectral efficiency of wireless networks  via aggressive spatial multiplexing  \cite{Marzetta2010,Lu2014,Rusek2013,Larsson2014}.
 Massive MIMO has extensively been studied with OMA, and it is known that, with the prevalent linear processing at the BS, the  best spectral efficiency of massive MIMO-OMA systems is obtained  in \textit{underloaded} systems. Specifically, for the  maximum ratio (MR) combining and
 zero-forcing (ZF) combining, the highest spectral efficiency is achieved when the number of users is about 2 and 5 times less than the number of antennas \cite{bjornson2016massive}.
Therefore, massive MIMO-OMA  may not be able to support the \textit{overloaded} systems, i.e., when the number of users exceeds the number of antennas at the BS.
Massive MIMO-NOMA, on the other hand, has a great potential to overcome this limit and to support  massive connectivity requirements of the next-generation wireless networks while further improving the spectral efficiency of  NOMA-based systems \cite{Senel2017}.
We  further investigate these potentials in Section~\ref{sec:massive}.

\subsubsection{Millimeter wave  communications}  To fulfill extremely high data  rate requirements of next-generation wireless  networks, the communication at  millimeter wave (mmWave) bands (30\,GHz to 300\,GHz spectrum)  has been
  subject to intensive research during the past decade, and it  has been
  proven theoretically and experimentally to provide gigabit-per-second data rates
  due to huge available  bandwidths (e.g. up to 2\,GHz bandwidth in 60\,GHz) \cite{Rappaport2014,Heath2016}.
 However, since mmWave channels are sparse in spatial/angle domain,
 the number of simultaneous connections  at these very high frequencies has shown to be  limited. Coupled with mmWave massive MIMO, NOMA can  circumvent this limit. Section~\ref{sec:mm} details this by summarizing the state-of-the-art on  the coexistence  of NOMA and mmWave massive MIMO.

\subsubsection{Cooperative  communications}
Cooperative communications has great potential to improve wireless networks throughput
and is a  well-investigated  area of research.
The key idea of cooperation is to share resources  among multiple nodes in a network. With user-cooperation, e.g., sharing power and computation with certain nodes,
overall network performance can be improved Include these two references \cite{Sendonaris2003a,Sendonaris2003b}. Different relaying schemes, device-to-device communication, and
 multi-cell cooperative transmission are among the well-known operative scheme.
NOMA and cooperative communications are capable of mutually supporting each other. Particularly, relay-aided NOMA and multi-cell cooperation have recently attracted considerable attention, for their great promise in improving spectral efficiency and user fairness.
These schemes will be  discussed
in Section~\ref{sec:co}.

\subsubsection{Cognitive communications}
Cognitive radio (CR) networks exploit   spectrum sharing to improve spectral efficiency \cite{Haykin2005}. Overlay CR is to employ  unused  spectrum and is usually performed opportunistically but CRs can work concurrently with incumbent users. 
CRs sense spectrum, detect incumbent users, and efficiently allocate/use spectrum.
  Similar to  cooperative communications, cognitive communications promises  great potential for spectral efficiency and has been under investigation for more than two decades.
NOMA can be applied to CR networks to increase the number of users and further increase the spectral efficiency. On the other hand, the principle of underlay CR can be utilized to design the so-called CR-inspire NOMA.
The interplay between CR and NOMA  will be  discussed
in Section~\ref{sec:cog}.

\subsubsection{PHY security} Physical layer (PHY) security  is a means of complementing  higher-layer
cryptographic security measures in   wireless  networks.
Unlike cryptographic approaches, PHY security approaches can
guarantee information secrecy regardless of an eavesdropper's
computational capability.
PHY security techniques  exploit  the physical characteristics of the
wireless communication channel, e.g.,  \textit{noise}, \textit{fading}, and  \textit{interference},
to guarantee secure communication directly at the physical layer.
These approaches aim at degrading the quality of signal reception at
eavesdroppers compared to  the
main channel and thereby preventing them from decoding the confidential information from
the intercepted signals \cite{mukherjee2014principles, shiu2011physical,  chen2017survey,liu2017physical}.
NOMA-based systems are susceptible to   eavesdroppers and  can benefit from  PHY security  as we elaborate this
in Section~\ref{sec:pls}.

\subsubsection{Energy harvesting}

Energy harvesting communications offer the promise of providing  energy self-sufficient and self-sustaining means of communications; a step toward realizing {\it green} communications \cite{Ulukus2015}. While energy is usually harvested from external natural sources, it could also be harvested from ambient radio frequency (RF) electromagnetic signals. The notions of simultaneous wireless and information power transfer (SWIPT) and wireless-powered communication networks (WPCN) are introduced and thoroughly studied in recent literature for that purpose. In SWIPT, energy is provided {\it along the way} during information transmission, in which users either employ power switching or time switching techniques to harvest energy (partially) from the transmitted signals. While in WPCN, energy is transferred wirelessly toward intended users so that they use it, mainly, to communicate back to the energy-providing sources. It is clear that such notions offer great potential to enhance different NOMA performance metrics, such as energy efficiency, achievable rates, and outage probabilities. We elaborate on this further in Section~\ref{sec:eh}.

\subsubsection{Visible light communications (VLC)}

VLC presents solutions to spectrum congestion and scarcity issues, especially in indoor environments, through shifting transmission frequencies from conventional RF ranges to the visible light range \cite{Pathak2015}. As such, transmit antennas become light emitting diodes (LEDs) and receive antennas become photodetectors (PDs). VLC is the enabling technology behind realizing light fidelity (LiFi) networks, and lots of research are currently being conducted to improve the performance of VLC. It is therefore amenable to combine VLC with NOMA techniques in multiuser scenarios to provide an efficient way to handle the available resources. We discuss this idea further in Section~\ref{sec:vlc}.

\subsubsection{Mobile edge computing (MEC)}

 MEC has emerged as a means of providing remote computation for mobile devices in 5G wireless systems and  is driven by the increasing demand for traffic volume and computation raised by the emerging compute-intensive applications, e.g., virtual reality and interactive gaming \cite{PMMECSurvey2017}. Due to their low computing capability,  mobile devices can offload their tasks to the BSs equipped with an MEC server. The offloaded tasks will be executed by the MEC server, and the result will be downloaded to the mobile devices after computing. In the MEC system, there are two offloading scenarios, i.e., \emph{partial} offloading and \emph{binary} offloading. In the partial offloading, the task can be partitioned into two main parts, i.e., offloading part and local computing part. In the binary offloading scheme, the task cannot be partitioned. It means each task will be either offloaded to the MEC server for remote execution or locally computed by the mobile devices.

\subsection{Organization and Existing Survey Papers}
As summarized in  Table.~\ref{table:org}, the remainder of this paper is organized as follows. In
Sections~\ref{sec:massive} and \ref{sec:mm}, the coexistence of NOMA with massive MIMO and mmWave massive MIMO  is described, respectively.
Sections~\ref{sec:co} and \ref{sec:cog}
elaborate on interaction of NOMA with cooperative
and cognitive communications, respectively.
PHY security approaches enabling the exchange of confidential messages over NOMA-based
networks in the presence of in network and external eavesdroppers are described in Section~\ref{sec:pls}. The state-of-the-art in the burgeoning area of energy harvesting NOMA are detailed in Section~\ref{sec:eh}.
Section~\ref{sec:vlc} discusses combining NOMA with VLC and the effect of that on more efficient resource management. Section~\ref{sec:mec} highlights possible application of NOMA to the emerging area of mobile edge computing. Finally, Section~\ref{sec:other} briefly lists other technologies benefiting from NOMA as well as applications of machine learning and deep learning in solving different problems in NOMA, including clustering and power allocation. Section~\ref{sec:conc} concludes the paper.

There exist a number of interesting survey papers in this topic	\cite{wei2016survey,islam2016power,ding2017survey,Liu2017Proceeding}. 
 This survey paper is different from the existing ones in that we are specifically focusing on  the  contributions that combine NOMA with the emerging technologies listed in Section~\ref{sec:tech} and emphasis 
 how NOMA and these prominent technologies interplay with,  benefit from, and shape the future of research in wireless networks. Further, this paper includes  recent progresses  reported in recently-published  works that have not been  included in previous surveys.


\section{Coexistence of NOMA and sub-6 GHz massive MIMO}
\label{sec:massive}
Massive MIMO systems operating in sub-6\,GHz frequency bands primarily rely on substantial  spatial multiplexing gains and favorable propagation characteristics rendered by very large antenna arrays to simultaneously serve  many user nodes in the same time-frequency resource element \cite{Marzetta2010}.    Sub-6 GHz massive MIMO has been shown to provide unprecedented spectral/energy efficiency gains \cite{Lu2014,Rusek2013,Larsson2014},  and  it has already been deployed by commercial carriers such as Sprint in the United States \cite{Sprint}.    Despite these benefits, massive MIMO with OMA may not be able to support the overloaded case in which the number of user nodes exceeds the number of RF chains at the BS. Hence, sub-6 GHz massive MIMO with OMA may not be able to support the massive connectivity requirement of the next-generation wireless standards.
To this end,  joint benefits of  NOMA and sub-6 GHz massive MIMO can be efficiently leveraged to satisfy future demands for  massive wireless  connectivity with high data rates and low-latency
\cite{Chen2018a,Cheng2018b,Kudathanthirige2019,Senel2017,Chen2018c,Xu2017,Liu2018,LiuX2017,Ma2017,Li2018b,Li2019b,Kusaladharma2019a,Sharath2019a,Silva2019,Aravinda2019,Li2018a}.

                \begin{figure}[!t]
                                                \centering
                \fontsize{9}{5}\selectfont
  \includegraphics[width=0.4\textwidth]{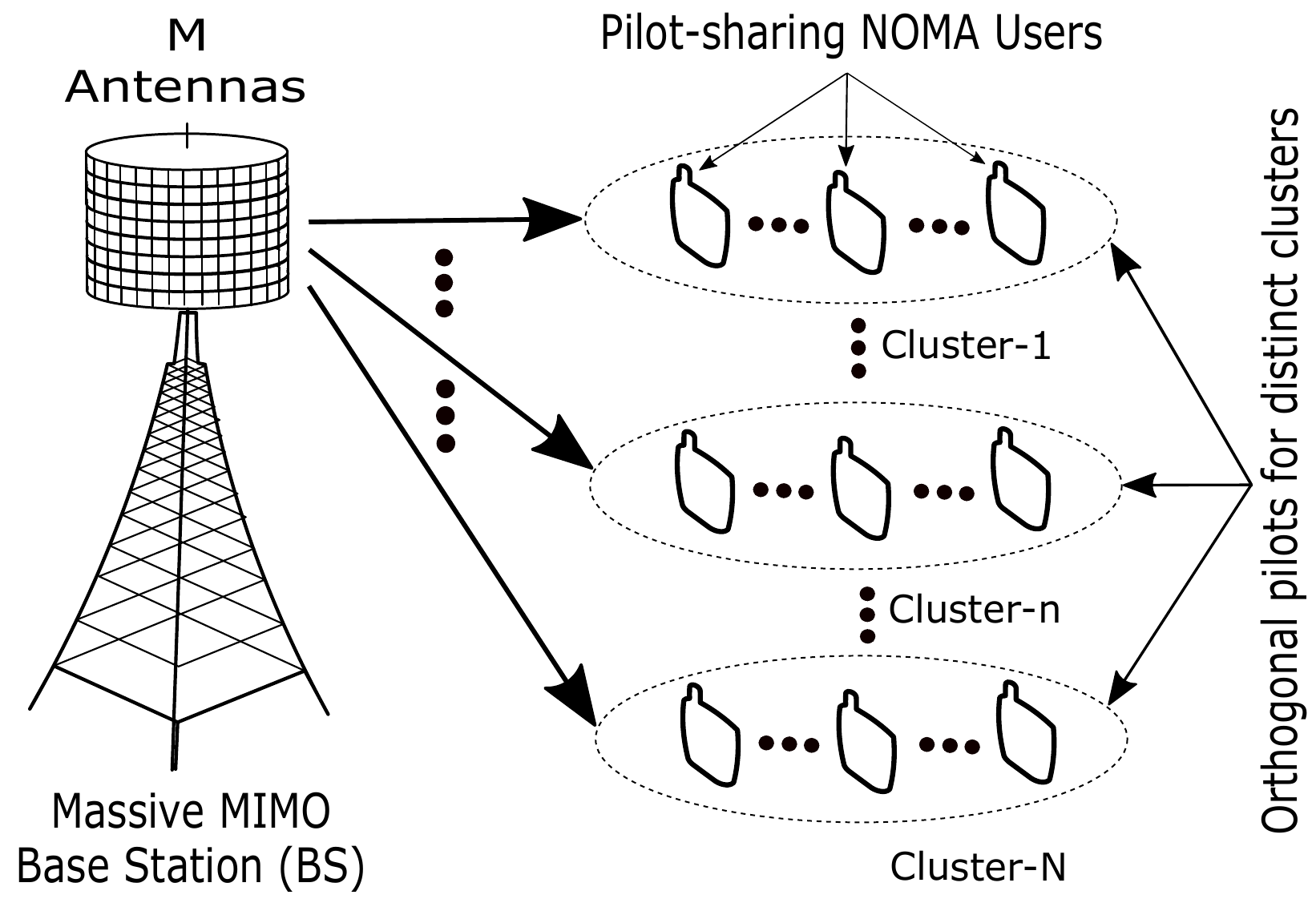}\vspace{-2mm}
                \caption{A system model for massive MIMO-NOMA with non-orthogonal pilot allocation within clusters \cite{Cheng2018b,Chen2018a}.} \label{fig:MMIMO_NOMA_System_Model}
                \vspace{-0mm}
                \end{figure}

  		\begin{table} [!h]
  			\caption {Summary of technical contributions on NOMA-enabled massive MIMO operating in sub-6\,GHz} 
  			\begin{center}{ 
  					\begin{tabular}{|m{0.5in}|m{ 2.2 in}|@{}m{0cm}@{}}
  						\hline 
  						\textbf{References}  & \textbf{Technical contribution} &\\ [1ex]
  						\hline\hline
  						\cite{Cheng2018b,Senel2017}  & Impact of estimated CSI on the performance bounds of single-cell   systems   with linear precoders  &\\ [1ex]
  						\hline 
  						\cite{Chen2018a,Kudathanthirige2019,Ma2017} & Pilot allocation strategies for multi/single-cell   systems    &\\ [1ex]
  						\hline
  						\cite{Sharath2019a} & Performance bounds for uplink superposition coded NOMA transmissions  &\\ [1ex]
  						\hline	
  						\cite{Chen2018c,Kudathanthirige2019} & User pairing/grouping and scheduling techniques and underlying performance bounds &\\ [1ex]
  						\hline  
  						\cite{Xu2017,Liu2018} & Iterative signal decoding algorithms for superposition-coded transmissions &\\ [1ex]
  						\hline  
  						\cite{Kudathanthirige2019,LiuX2017} & Transmit power allocation techniques and  performance comparisons  &\\ [1ex]
  						\hline    
  						\cite{Li2018b,Li2019b,Kusaladharma2019a} & Effects of distributed massive antenna systems on the performance bounds  &\\ [1ex]
  						\hline    
  						\cite{Zhang2017a,LiuX2017,Chen2018_Relay,Li2018c,Silva2019} & Integration of relaying techniques and performance analysis &\\ [1ex]
  						\hline    
  						\cite{Li2018a} & Integration with underlay spectrum sharing techniques &\\ [1ex]
  						\hline		
  					\end{tabular}\label{table:table_1}}
  			\end{center}
  	
  		\end{table}

	\subsection{Design Insights and Implications }
  	
  	In this subsection, we summarize   key design insights, implications of   practical transmission impairments, and  concluding remarks for sub-6 GHz massive MIMO-NOMA systems.

The importance of  acquisition of accurate CSI for integrating NOMA into massive MIMO has been elaborated in \cite{Cheng2018b,Chen2018a,Senel2017,Kudathanthirige2019}. The performance metrics derived by assuming the genie-aided perfect CSI can be misleading in a practical massive MIMO set-up because they always overestimate the system performance. To this end, in \cite{Cheng2018b,Chen2018a}, it has been revealed that in an overloaded NOMA system, the achievable performance is always detrimentally affected by  residual interference caused by  intra/inter-cluster pilot contamination even in the asymptotic BS antenna regime.

  	In \cite{Cheng2018b}, the performance of NOMA has been compared against the    conventional multi-user  spatial multiplexing with OMA for a training-based sub-6 GHz massive MIMO system by adopting a non-orthogonal pilot allocation (see Fig. \ref{fig:MMIMO_NOMA_System_Model}).
  	Reference \cite{Chen2018a} extends the system model and performance analysis  of \cite{Cheng2018b} to facilitate multiple cells.
  	It has been shown in  \cite{Cheng2018b} that NOMA outperforms OMA in terms of the achievable sum rate only when all user nodes are provided  with highly accurate    estimated CSI through beamforming downlink pilots   and with no inter-cluster interference. The performance gains of NOMA over OMA can be boosted  when there exist distinct path-losses of the channels pertaining to  user nodes located within the same cluster.
  	Moreover, \cite{Cheng2018b} advocates OMA with multiuser spatial multiplexing instead of NOMA when there is more than one NOMA user-cluster. It also  reveals the importance of accurate estimated CSI at both the BS and user nodes through uplink and downlink pilots transmissions, respectively.
  	The aforementioned conclusions have been drawn in \cite{Cheng2018b} for maximal ratio transmission  based precoders at the BS.

  	In \cite{Senel2017}, the performance gains that NOMA can provide in a sub-6 GHz massive MIMO system
  	have been investigated.  The achievable rate analysis of  \cite{Senel2017} reveals that  massive MIMO-OMA with ZF precoding outperforms power-domain NOMA when the number of antennas at the BS significantly exceeds the number of user nodes ($M\gg K$). Nevertheless, it has also been shown that NOMA performs better than multi-user massive MIMO when  the number of BS antennas and total user count are approximately equal ($M \approx K$). Moreover, the performance of massive MIMO-NOMA for a non-line-of-sight (NLoS) independent and identically distributed (i.i.d.) Rayleigh fading channel model has been compared against   a line-of-sight (LoS) deterministic channel model. It reveals that  the performance gains of NOMA become more prominent for the latter  channel model than the former counterpart.

  	In \cite{Kudathanthirige2019}, the achievable performance gains of NOMA over OMA have been compared for a multi-cell  massive MIMO system operating over spatially-correlated fading channels. The paper proposes a novel user classification/clustering and pilot allocation schemes based on the channel covariance information. It has been shown  that the proposed pilot allocation in \cite{Kudathanthirige2019} outperforms the conventional non-orthogonal pilot allocation of   \cite{Chen2018a,Cheng2018b} and the orthogonal pilot allocation of \cite{Ma2017} in terms of the achievable user rates. Moreover,  \cite{Kudathanthirige2019} reveals that the asymptotic achievable rates are no longer limited by the residual intra/inter-cell/cluster interference for the underloaded NOMA case in which the number of BS antennas exceeds the number of user nodes. Nonetheless, the performance gains will be degraded by intra-cluster pilot contamination when the covariance matrices of multiple user nodes located within the same cluster are linearly dependent in the asymptotic BS antenna regime. However, this condition rarely occurs in practice with sub-6 GHz frequency band as the covariance matrices will more likely to be asymptotically linearly independent \cite{Bjornson2018} when the number of antennas at the BS grows without a bound.
  	
  	The uplink specific transmission designs and performance analysis for NOMA-aided sub-6 GHz massive MIMO have been investigated in \cite{Sharath2019a}. Upon deriving the achievable uplink rates under
  	estimated uplink CSI at the BS, it has been shown in \cite{Sharath2019a} that proper transmit power control at the user nodes is an essential  design aspect to boost the achievable uplink rates of massive MIMO-NOMA. When no downlink pilots are transmitted, the user nodes must rely on statistical channel knowledge to implement transmit power control. To this end, \cite{Sharath2019a} proposes a  max-min optimal uplink transmit power control policy based on channel statistics to guarantee user-fairness in the presence of near-far effects in the uplink massive MIMO channels.

\subsection{Applications of   Massive MIMO-NOMA in Relay Networks}

 In \cite{Zhang2017a,LiuX2017,Chen2018_Relay,Li2019b,Silva2019}, the coexistence of sub-6 GHz massive MIMO-NOMA with relay networks has been investigated. By invoking the deterministic-equivalent techniques from random matrix theory, reference \cite{Zhang2017a} investigates the achievable rate bounds for massive MIMO-NOMA relay networks.
It has been shown  that the achievable sum rate increases linearly with the number of admitted relayed-users, while  the ratio between the transmit antenna count and  the relay count plays a key role in boosting the system-wide performance metrics. Reference  \cite{LiuX2017} reveals that   efficient three dimensional (3D) resource allocation techniques can further improve the performance of  NOMA-aided massive MIMO relaying.
In \cite{Chen2018_Relay},  the effects of system parameters on the achievable spectral efficiency gains of massive MIMO-NOMA relaying have been investigated for  three cases, namely (i) a large number of BS antennas, (ii) a large number of relay nodes, and (iii) a high level of BS transmit power. Moreover,  \cite{Chen2018_Relay} reveals the importance of efficient  transmit power optimization  at the BS and  relay nodes. In \cite{Li2018c}, the detrimental impact of practical transmission impairments such as the channel estimation errors, pilot contamination, imperfect SIC, intra/inter-cluster interference has been quantified for the relay-aided massive MIMO-NOMA downlink. In \cite{Silva2019}, for a $K$-user massive MIMO multi-way relaying, a novel NOMA transmission strategy, which can reduce the number of channel-uses to just two from
  $\left \lceil{{(K-1)}/{2}}\right \rceil +1$
  in the current state-of-the-art \cite{Ho2018},  has been investigated.

 \subsection{Massive MIMO-NOMA with Distributed Transmissions}

The feasibility of massive-scale distributed transmission to boost the performance of NOMA has been investigated in  \cite{Li2018b,Li2019b,Kusaladharma2019a}. To this end, in \cite{Li2018b}, the achievable downlink rates of a NOMA-aided cell-free massive MIMO system have been derived  in the presence of beamforming uncertainty and residual interference due to erroneous channel estimation and imperfect SIC operations.   Reference \cite{Li2019b} extends the single-antenna access-points (APs) of \cite{Li2018b} to  support multi-antenna APs in an attempt to leverage the benefits of   distributed multi-antenna transmissions
to boost the achievable rate performance of the  NOMA downlink. Moreover,   \cite{Li2018a} implements a cell-free  version of the underlay spectrum-sharing  massive MIMO-NOMA of  \cite{Aravinda2019} to ensure that   user-centric distributed transmissions improve the performance of secondary underlay spectrum-sharing without hindering the primary system performance gains. A summary of main contributions on massive MIMO-NOMA operating in sub-6\,GHz is listed in Table.~\ref{table:table_1}.


\section{Coexistence of NOMA and mmWave  massive MIMO}
\label{sec:mm}

The feasibility of supporting  gigabits-per-second data rates by wireless communications at  mmWave bands (30\,GHz to 300\,GHz)   has been verified both theoretically and experimentally \cite{Rappaport2014,Heath2016,Rangan2014,Rappaport2013}. The large   path-losses encountered at these mmWave frequencies can be compensated by leveraging the unprecedentedly high array gains that can be obtained by densely packing a massive number of antennas in a very-small area thanks to  much smaller wavelengths of mmWaves  \cite{Rappaport2014,Heath2016,Rangan2014,Rappaport2013}. In order to make massive MIMO at mmWaves practically feasible, hybrid beamforming, which cascades  a high-dimensional analog-precoder with a low-dimensional digital precoder,   is preferred over  full-dimensional digital beamforming \cite{Molisch2015,Sudarshan2006,Han2015,Sohrabi2016}. Due to the sparsity of mmWave channels in spatial/angle domain, the number of simultaneous connections that can be by virtue of  massive MIMO operating at these very high frequencies has shown to be  limited \cite{Wang2017,Wei2018}. To circumvent this issue, NOMA can be coupled with mmWave massive MIMO to drastically increase the number of simultaneously served user nodes in the same time-frequency resource element \cite{Zhang2017_mmWave,Ding2017a,Ding2017b,Mojtaba2018,Wang2017,Wei2018,almasi2019lens}. 
A list of important contributions of mmWave massive MIMO-NOMA can be found in Table.~\ref{table:table_2}.

\begin{table} [!t]
		\caption {Summary of technical contributions on NOMA-enabled massive MIMO operating at mmWaves} 
	\begin{center}{ 
			\begin{tabular}{|m{0.5in}|m{ 2.2 in}|@{}m{0cm}@{}}
				\hline 
				\textbf{References}  & \textbf{Technical contribution} &\\ [1ex]
				\hline \hline
				\cite{Zhang2017_mmWave}  & Achievable rates and capacity bounds via deterministic equivalent technique   &\\ [1ex]
				\hline 
				\cite{Ding2017a} & Limited feedback technique and corresponding performance comparisons    &\\ [1ex]
				\hline
				\cite{Ding2017b} & Finite resolution analog beamforming techniques and performance bounds  &\\ [1ex]
				\hline	
				\cite{Mojtaba2018} & Impact of beam misalignment in hybrid beamforming &\\ [1ex]
				\hline  
				\cite{Wang2017} & Integration of beamspace techniques with lens antenna arrays &\\ [1ex]
				\hline  
				\cite{Wei2019,Wei2019a} & Multi-beam techniques and beamwidth control algorithms   &\\ [1ex]
				\hline    	
			\end{tabular}\label{table:table_2}}
	\end{center}
\end{table}

\subsection{Design Insights and Implications}

In this subsection, we summarize notable contributions to the development of mmWave massive MIMO-NOMA systems.   Moreover, the corresponding key design insights, implications of   practical transmission impairments, and  conclusions are summarized.

The achievable rate bounds of mmWave massive MIMO-NOMA systems have been derived in \cite{Zhang2017_mmWave}, and thereby, it has been concluded that a combination of  massive MIMO with NOMA and mmWaves can provide very large 	spectral efficiency gains.
The achievable rates have been computed for two signal-to-noise ratio (SNR) regimes. By invoking the deterministic equivalent technique  with the Stieltjes-Shannon transform, the rate bounds have been established for the noise-dominated low SNR regime. In the interference-limited high SINR regime, the corresponding rate bounds have been derived by using the   channel statistics and   eigenvalue distributions. These achievable rate bounds may be useful as benchmarks for comparison purposes of practically-viable  mmWave massive MIMO-NOMA designs with estimated CSI, imperfect SIC, and residual inter/intra-cell interference.

 Reference \cite{Ding2017a} proposes a  low-feedback NOMA design, which  decomposes the massive MIMO-NOMA channel into a set of SISO-NOMA channels and thereby significantly reduces the computational complexity. By invoking perfect user-ordering and one-bit feedback, a performance analysis framework  has been developed. Thereby,  the proposed system model in \cite{Ding2017a}  strikes a balance between the achievable performance gains and the implementation/computational complexity.

  Full CSI acquisition and feedback may be prohibitively complicated for analog precoding based mmWave massive MIMO \cite{Rappaport2014,Molisch2015}. To this end, finite-resolution analog precoder designs have been adopted to substantially reduce the hardware cost \cite{Rappaport2014}. Consequently, the utilization of such finite-resolution precoders results in mismatches of beam alignments, which in turn yield received power leakages. While  imperfect beam alignment  is detrimental in OMA-based mmWave massive MIMO,   it can  be beneficial in sharing a beam with multiple user nodes clustered together and serving those users  by virtue of  NOMA supposition-coded transmissions.   Having been inspired by this observation, in \cite{Ding2017b}, NOMA has been used to circumvent the loss of  degrees-of-freedom  in finite resolution analog beamforming by  serving a set of user nodes via analog beam-sharing.

 In \cite{Mojtaba2018}, the detrimental impact of beam misalignment in mmWave massive MIMO-NOMA systems with hybrid beamforming has been investigated. The optimal analog and digital precoders have been designed based on maximizing the sum rate. To this end, a lower bound of the achievable sum rate has been derived in closed-form by assuming  perfectly aligned LoS channels.
 The impact of  misaligned  LoS  or NLoS  channels has been captured by invoking a carefully designed beam  misalignment  factor. Then, a lower bound of the achievable user rates under beam misalignment has been computed.  Finally, an upper bound of the rate gap between the perfectly aligned and misaligned beams  has been derived. Thereby, it has been concluded in  \cite{Mojtaba2018} that the achievable rates can be severely affected when imperfect designs of analog/digital precoders yield beam misalignments in mmWave massive MIMO-NOMA systems.

  NOMA has been integrated into mmWave  MIMO with lens antenna arrays  in \cite{Wang2017} and  \cite{almasi2019lens}.
 By adopting the proposed methods, the number of users that can be served simultaneously can exceed the number of RF chains.  By invoking a lens antenna array, the conventional spatial MIMO channel can be transformed into  beamspace domain via a discrete Fourier transform  \cite{Brady2013}.  Then, the achievable rates have been derived, and the corresponding analysis has been  used to design precoders to reduce inter-beam interference and to formulate   transmit power control algorithms.   
 In order to maximize the achievable sum rate, a dynamic power allocation scheme has been formulated to minimize intra/inter-beam interference.  This dynamic power allocation problem has been solved via a low-complexity iterative optimization algorithm. By exploiting   efficient beam selection algorithms for beamspace-domain, it has been revealed in \cite{Wang2017} that the achievable spectral and energy efficiency gains can be considerably boosted by integrating NOMA into the beamspace mmWave MIMO.

 In all aforementioned related literature on mmWave massive MIMO NOMA \cite{Zhang2017_mmWave,Ding2017a,Ding2017b,Mojtaba2018,Wang2017}, each clusters of user nodes  having similar spatial signatures have been served by a superposition-coded signal through a  single beam. However, reference  \cite{Wei2019} reveals that only a few user nodes can be served via a single beam once mmWave communications is adopted due to extremely narrow beamwidths. To circumvent this issue, the concept of multi-beam mmWave massive MIMO has been proposed in \cite{Wei2019}. To this end,  multiple beams can be used to serve a single-cluster through efficient antenna partitioning algorithms. Reference   \cite{Wei2019} concludes that multi-beam NOMA transmissions can significantly boost the achievable rates at the user nodes by mitigating the adverse effects of beam misalignment exhibited in single-beam mmWave massive MIMO NOMA systems.  Moreover, in \cite{Wei2019a}, two  beamwidth  control  techniques have been proposed for multi-beam mmWave massive MIMO NOMA via   the   conventional   beamforming   and   the   Dolph-Chebyshev beamforming. By computing the main  lobe  power  losses incurred by the proposed beamwidth control, reference \cite{Wei2019a}  proposes    an  effective  analog  beamformer, a novel resource allocation scheme,  and a NOMA user grouping algorithm based on the coalition formation game  theory to boost the overall system performance.

Conclusions and Future Research Directions }
In the related prior research on NOMA-enabled massive MIMO \cite{Chen2018a,Cheng2018b,Kudathanthirige2019,Senel2017,Chen2018c,Xu2017,Liu2018,LiuX2017,Ma2017,Li2018b,Li2019b,Kusaladharma2019a,Sharath2019a,Li2018a,Zhang2017a,Chen2018_Relay,Li2018c,Silva2019,Zhang2017_mmWave,Ding2017a,Ding2017b,Mojtaba2018,Wang2017,Wei2019,Wei2019a}, the spatial domain has solely been exploited in designing pilot allocations, user groupings/pairings, and related signal processing techniques. 
Exploitation of the angle domain with array signal processing techniques has recently gained much attention \cite{Zhao2017,Lin2017,Fan2017}.  To this end, angular models of the massive MIMO NOMA channels can be leveraged to design angle information aided pilot allocation and channel estimation, beamforming and power allocation and interference mitigation  techniques by virtue of theory of array signal processing \cite{Van_Trees_2002}.      

Deep learning techniques by virtue of artificial neural networks can be  a useful tool in enhancing the efficiency of  data/model-driven transmitter/receiver designs  for  NOMA-aided massive MIMO. In particular, the fundamental trade-offs among system parameters involved in channel estimation, transmit power allocation and  iterative/SIC decoding can be optimized by virtue of tools in deep learning.


\section{Coexistence of NOMA and Cooperative Communications}
\label{sec:co}

Broadly speaking,  research contributions on the coexistence of NOMA and cooperative communications can be divided  in three categories, namely, cooperative NOMA, relay-aided NOMA communications, and  multi-cell NOMA  cooperative transmission, as detailed in the following. 

\subsection{Cooperative NOMA}

The basic principle of cooperative NOMA is to invoke one NOMA user as a relay, as proposed in \cite{ding2015cooperative}. More specifically, the transmission process of cooperative NOMA consists of two time-slots. The first time-slot, the BS broadcasts the superposed messages to two NOMA users. At the second time-slot, the user with good channel condition acts as a decode-and-forward (DF) user relay to forward the decoded messages to the user with poor channel conditions. As a consequence, the reliability of weak user can be improved. Compared to conventional NOMA, the key advantages of cooperative NOMA are that this scheme is capable of achieving low system redundancy, better fairness, and higher diversity gain, which has been summarized in \cite{Liu2017Proceeding}.  To elaborate further, an energy efficient energy harvesting cooperative NOMA protocol was proposed in \cite{Liu2016JSAC1}, with invoking a stochastic geometry model. Three user selection schemes are proposed based on the user distances from the base station. Although cooperative NOMA is capable of enhancing the performance of poor user's, it costs one extra slot to transmit information. Full duplex relay technique is a possible technique for solving this issue. The applications of full-duplex relay on cooperative NOMA have been considered in \cite{Zhong2016CL,Zhang2017TVT,Yue2018TCOM,Elbamby2017JSAC}. To further enhance the performance of cooperative NOMA, a two-stage relay selection scheme was proposed  in \cite{Ding2016WCL} to minimize the outage probability among possible relay selection policies. Considering the spatial effects, the  relay selection  was investigated in \cite{Yue2018TCOM}, where the spatial random distributed relays are capable of switching between half-duplex mode and full-duplex mode.

\begin{table*}[ht!]
	\caption{Summary of existing works on NOMA and Cooperative Communications}
		\begin{center}{ 
	\begin{tabular}{|p{3.3cm}|p{2cm}|p{8.5cm}|p{1.7cm}|}
		\hline 
		\textbf{Category} & \textbf{Transmission} & \textbf{Advantages/Contributions}&\textbf{References}\\
		\hline \hline		Cooperative NOMA & Downlink & Lower system redundancy, enhanced fairness and diversity gain  & \cite{ding2015cooperative,Liu2016JSAC1,Zhong2016CL,Zhang2017TVT,Yue2018TCOM,Elbamby2017JSAC,Ding2016WCL} \\
		\hline
		Relay-aided NOMA&Downlink/Uplink& Extend the network coverage & \cite{Men2015CL,Shin2017SPL,Kim2015CoMP,Yue2018twoway} \\
		\hline
		Multi-cell cooperate NOMA &Downlink/Uplink&  Enhance the performance of cell-edge users and increase spectral efficiency  & \cite{shin2017non,Shin2017CL,Liu2017JSAC,Zhou2018Network,Ali2018WCM,Liu2018MIMONOMA,shin2016number,Gu2018CL,Qin2018Mag}  \\
		\hline		
	\end{tabular}
	\label{table:cop}}
	\end{center}

\end{table*}

\subsection{Relay-aided NOMA}
In this context, several relay-aided NOMA communications schemes are proposed for single-cell NOMA networks~\cite{Men2015CL,Shin2017SPL,Kim2015CoMP,Yue2018twoway}. Notably, an amplify-and-forward (AF)  multi-antenna relay-aided NOMA downlink network was investigated with obtaining the outage performance in \cite{Men2015CL}. Regarding the uplink case, a novel relay-aided NOMA scheme was proposed in \cite{Shin2017SPL} for multi-cell scenarios, where an Alamouti structure was applied. In \cite{Kim2015CoMP}, a novel coordinated direct and relay-aided transmission scheme was proposed, where the BS communicates with the near user directly while communicates with the far user with the aid of a relay.  As a further advance, in \cite{Yue2018twoway}, the authors proposed a novel two-way relay NOMA scheme, where a pair of NOMA users can exchange their information with the aid of decode-and-forward relay. Relay-aided physical layer security, which  is of significant importance in NOMA, will be discussed in  Section~\ref{sec:pls}. 

\subsection{Multi-cell NOMA Cooperative Transmission}

As research contributions in the context of single cell-NOMA have been well researched, researchers recently focus their attention on multi-cell NOMA, such as network NOMA \cite{Dai2015NOMA}, coordinated multi-point cooperative (CoMP) transmission \cite{shin2017non}, NOMA in heterogeneous networks (HetNets)\textit{} \cite{Liu2017JSAC}, NOMA in cloud radio access networks (C-RAN)~\cite{Zhou2018Network}, etc. For multi-cell NOMA scenarios, one major concern is  how to enhance the performance of cell-edge NOMA users, especially for the downlink case. This is because, typically, the far users (cell-edge user) is  not well-served. Another key issue of multi-cell NOMA is to handle the complicated intra/inter cell interference. CoMP is a promising technique to solve the aforementioned issue, by enabling multiple BSs to carry out coordinated beamforming for enhancing the performance of cell-edge user \cite{Ali2018WCM}. As multiple antenna techniques are significant for NOMA by bringing additional gain in spatial domain \cite{Liu2018MIMONOMA},  two novel coordinated beamforming approaches were proposed in multi-cell MIMO-NOMA systems for decreasing the inter-cell interference in \cite{Shin2017CL}. 
A majorization-minimization-based  beamforming is shown to
outperform ZF-based beamforming in secure  rate optimization  of a two-user MIMO-NOMA network in \cite{jiang2017secure}. Moreover, in \cite{Gu2018CL}, the outage performance was investigated in downlink C-RAN NOMA networks, where stochastic geometry was used for modeling the locations of BSs and NOMA users. It is worth to point out that there are still several open issues for multi-cell NOMA cooperative transmission \cite{shin2017non}. For example, compared to single-cell cooperative NOMA, the SIC decoding order is required to be reconsidered in multi-cell cooperate NOMA as the near user is not necessary to be the user with good channel quality \cite{Qin2018Mag}. Moreover, the power control/allocation among users also play a pivotal role for multi-cell NOMA cooperate transmission (especially for HetNets NOMA scenarios \cite{Liu2017JSAC}) as inappropriate power allocation will result in increased energy consumption hence degrade the system performance.  Other issues such as error propagation, hardware complexity for decoding also hinder the implementation of SIC in multi-cell NOMA and need further investigation.
A high-level summary of existing works on NOMA with cooperative communication is listed in Table.~\ref{table:cop}.

\subsection{Discussions and Outlook}
There are still several open issues for multi-cell NOMA cooperate transmission \cite{shin2017non}. For example,  SIC decoding order is required to be reconsidered in multi-cell cooperate NOMA as the near user is not necessary to be the user with good channel quality \cite{Qin2018Mag}. Moreover, the power control/allocation among users also plays a pivotal role for multi-cell NOMA cooperate transmission (especially for HetNets NOMA scenarios \cite{Liu2017JSAC}) as inappropriate power allocation can increase energy consumption and hence degrade the system performance.  Other issues such as error propagation, hardware complexity for decoding also hinder the implementation of SIC in multi-cell NOMA \cite{vaezi2018non,shin2017non}. More research contributions are required to make multi-cell NOMA practical.


\section{Interplay between NOMA and Cognitive Radio Networks}
\label{sec:cog}

Proposed by Mitola in 2000 \cite{mitola2000cognitive}, the concept of cognitive radio (CR), which allows the unlicensed secondary users to opportunistically access the licensed primary users' spectrum, has received significant attention both in industry and academia. CR can be broadly categorized as three paradigms,  overlay \cite{goldsmith09,vaezi2012capacity}, underlay \cite{Liu2015WPT} and interweave \cite{QINTCOM2017,QINSPM2018} (see Table 8 of \cite{Liu2017Proceeding} for the detailed comparison of the three schemes).  Given the fact that both NOMA and CR techniques are capable of enhancing the spectral efficiency, a natural question arises: Can NOMA and CR coexist to achieve an improved spectral efficiency? The investigation of co-existence of NOMA and CR can be summarized from two aspects: 1) The application of NOMA in CR networks; 2) The cognitive radio inspire NOMA, which will be detailed in the following.

\subsection{NOMA in Cognitive Radio Networks}
\begin{figure}
\centering
\includegraphics[width=0.47\textwidth]{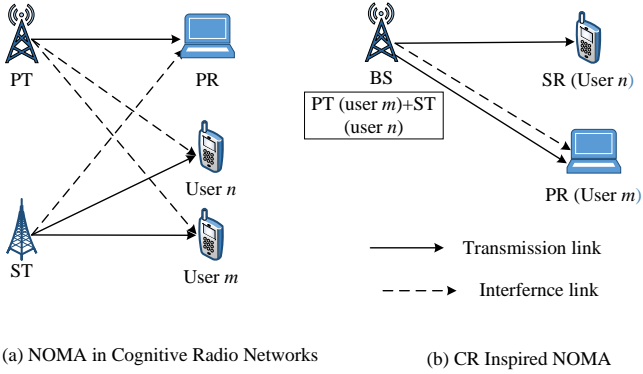}
\caption{Interplay between NOMA and cognitive radio networks.}
\label{fig:CRNOMA}
\end{figure}

The key idea of applying NOMA in CR networks is to allow the secondary transmitter (ST) to communicate with multiple secondary NOMA users as long as the interference power constraint at the primary user (PUs) is satisfied. Fig. \ref{fig:CRNOMA}~(a) illustrates a two-user case NOMA in CR networks, where ST is capable of communicating with User~$n$ and User~$m$ simultaneously. In \cite{Liu2016TVT}, the authors first proposed to apply NOMA in large-scale CR networks with the aid of invoking stochastic geometry model. A general NOMA-CR scenario is considered, where one ST is capable of communicating with multiple NOMA SUs. As a further advance for the application of NOMA in CR networks, in \cite{Lv2017TCOM}, the authors proposed a cooperative NOMA-CR scheme for enabling multiple SUs to serve as relays to aid transmission. Moreover, for further enhancing the spectral efficiency,  two novel NOMA-CR user scheduling schemes were considered in \cite{Lv2018TVT} to ensure NOMA users to be scheduled either efficiently or fairly.

\subsection{Cognitive Radio-inspired NOMA}
In this subsection, the CR inspired NOMA is introduced. As proposed in \cite{ding2014pairing}, the key concept of CR inspired NOMA is essentially a novel power allocation scheme. As shown in Fig.~\ref{fig:CRNOMA}~(b), we can consider the BS as a combination of the primary transmitter (PT) and ST, which transmits the superposition-coded signals. The rationale behind CR inspired NOMA is to ensure the quality-of-service (QoS) of the weak user (User~$m$) by limiting the power allocated to the strong user (User~$n$). In this case, we can still investigate this scheme with the aid of the key feature of classic underlay CR. In \cite{ding2014pairing}, the impact of user pairing on the performance of both fixed power allocation and CR inspired NOMA has been investigated. A more general power allocation scheme for guaranteeing the QoS in both downlink and uplink NOMA systems has been considered in \cite{Yang2016TWC}. By applying such a scheme, it is more flexible to achieve the throughput/fairness tradeoff. Regarding the extension to MIMO scenarios, the authors in \cite{Ding2016MIMO} apply the CR inspired NOMA for solving the power allocation issue in MIMO-NOMA networks with the aid of signal alignment.

The aforementioned existing research contributions in the context of the interplay between CR and NOMA networks mainly focus on underlay CR. The research on  interweave NOMA-CR and overlay NOMA-CR is still in its infancy, and thus, research advancements on the corresponding open problems are expected in the future. It is also worth  pointing out that proper resource management, such as power allocation, user clustering, and user pairing can be considered to further enhance the system performance of NOMA-CR networks.

\subsection{Discussions and Outlook}
Existing research on NOMA-CR networks mainly focus on underlay CR. The works on interweave overlay  NOMA-CR are still in theirs infancy. Better resource management, such as power allocation and
	user clustering,  can be considered to further enhance the system performance of NOMA-CR networks. The use of machine learning, and deep learning, has attracted extensive research interest to enable  intelligent CR from both physical layer and resource allocation~\cite{QIN:WM:2019,liu2019machine}. There is still a long way  to realize intelligent NOMA-CR before we solve the following challenges:
\begin{itemize}
	\item NOMA-CR-enabled massive connectivity: 
 To support massive connectivity with low power consumption, low-power wide-area techniques, such as narrow-band IoT and LoRa, have been proposed to support short packet transmission~\cite{QIN:IOT:2019}. Since many  IoT devices are power constrained, it is worth  investigating NOMA-CR techniques with low power consumption to make them applicable to massive connectivity of IoT devices.
	\item Intelligent/distributed resource management: Resource allocation in NOMA-CR has been investigated extensively. important emerging  applications such as unmanned aerial vehicles (UAVs) and vehicle-to-everything (V2X)   pose stringent challenges on intelligent resource management. Existing resource allocation schemesare implemented in a centralized approach~\cite{Cui:JSTSP:2019} which may not suitable for dynamic UAV deployment as  UAVs are expected to make a decision locally \cite{Liu2019WCM}.
 A distributed spectrum sharing scheme has been proposed in \cite{liang2019spectrum} for the V2X  links. 
More efficient spectrum sharing schemes are expected for the NOMA-CR systems.
\end{itemize}

\section{NOMA and Physical Layer Security}
\label{sec:pls}

\begin{figure}[t]
	\center
	\includegraphics[scale=.99]{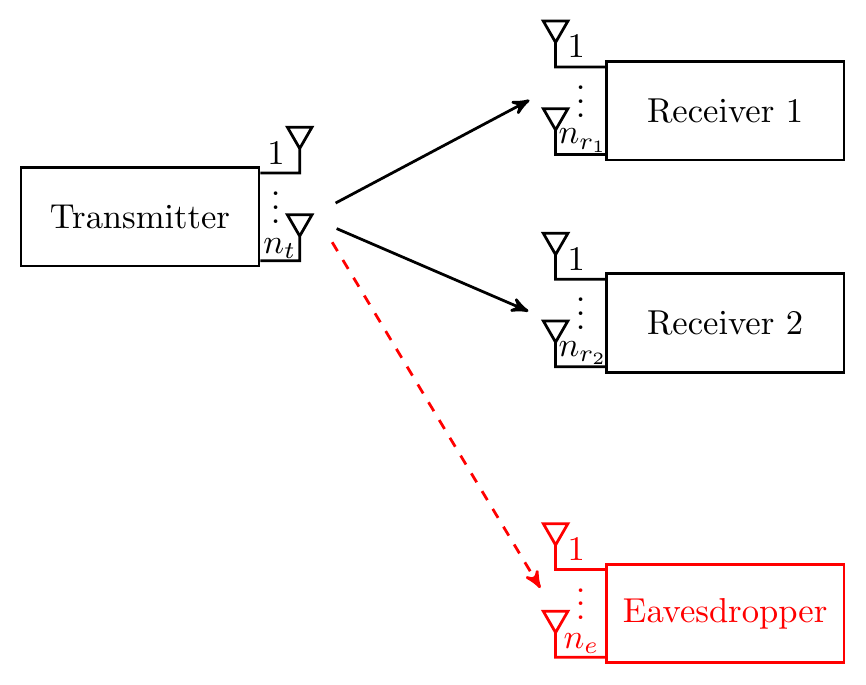}
	\caption{A two-user MIMO-NOMA with one external eavesdropper  \cite{vaezi2019noma}. All nodes are equipped with multiple antennas.\
	This basic model can be extended in several ways, see \cite{vaezi2019noma} for more details. Each NOMA user may also be seen as an (internal) eavesdropper to the other NOMA user. }
	\label{fig:pls} 
\end{figure}

Due to the broadcast nature of wireless
transmissions, securing transmitted data from potential external eavesdroppers and internal eavesdroppers (untrusted nodes in the
network) is a critical system design aspect that needs careful consideration.
PHY security
is a powerful tool to achieve the goal of a provably unbreakable, secure communications. PHY security
techniques exploit the physical aspects of communication channels between the nodes to introduce
security in wireless communication systems.
Today, several PHY security techniques are known to guarantee  secure communication
in the context of wiretap channel.
Some can guarantee security even if the
legitimate user's channel is worse than the eavesdropper's channel.
Most notably among them are \textit{artificial noise} (AN) transmission to confuse the eavesdropper \cite{goel2008guaranteeing},
various beamforming approaches  \cite{fakoorian2012optimal,shafiee2009towards,vaezi2017mimo,vaezi2017optimal},
transmit antenna selection \cite{ferdinand2013effects,yang2013transmit,alves2012performance},
cooperative jamming and relay-based PHY security systems \cite{dong2010improving,jeong2011optimal,li2011cooperative}.
 The above approaches are, in
 general, applicable to NOMA-based networks \cite{vaezi2019noma}, and many of them have already been investigated
 in this context, as described below.
A  two-user MIMO-NOMA with an external eavesdropper is shown in Fig.~\ref{fig:pls}.

 \begin{table*} [!t]
 	\caption {Summary of technical contributions on NOMA-related physical layer security} 
  	\begin{center}{ 
 			\begin{tabular}{|m{0.6in}|m{1.5in}|m{ 3 in}|@{}m{0cm}@{}}
 				\hline 
 				\textbf{References} & \textbf{System model}& \textbf{Technical contribution} &\\ [1ex]
 				\hline \hline
 				\cite{lv2018secure}  & Two-user MISO-NOMA & Secrecy outage probability  is evaluated
 				    \\ [1ex]
 				    	\hline
 				    \cite{zhou2018artificial}  &$K$-user MISO-NOMA cognitive radio network  & AN-aided
 				    cooperative jamming is used to improve
 				    the security of the primary user
 				    \\ [1ex]
 					\hline
 				\cite{zeng2019securing}  & Massive MIMO-NOMA &Ergodic secrecy rate is evaluated  \\ [1ex]
 					\hline
 				\cite{liu2017enhancing}  & Large scale SISO/MISO-NOMA &Secrecy outage probability is derived    \\ [1ex]
 					\hline
 				\cite{zheng2018secure}  & Two users, one trusted relay,  and  multiple
 				eavesdroppers & Jamming is transmitted to increase security and  achievable ergodic secrecy rates are derived    \\ [1ex]			
 				\hline 
 			\cite{arafa2018secure}  & Two-user SISO-NOMA with  external eavesdroppers  &Secrecy rates are derived with different trusted/untrusted relaying including cooperative jamming    \\ [1ex]
 				\hline
 					\cite{xiang2019physical}  & SISO NOMA-CR  where secondary users are eavesdroppers  &Connection outage probability, secrecy outage probability, and effective secrecy
 					throughput are derived for the primary users  \\ [1ex]
 				\hline  
 				\cite{yuan2019analysis}  & Cooperative SISO-NOMA with an  external eavesdropper  & Expressions for the ergodic secrecy rates are derived in the presence of an eavesdropper where he source transmits jamming signals  \\ [1ex]
 				\hline  
 				\cite{xiang2019secure}  & Two-user SISO-NOMA with an untrusted relay  &Effective
 				secrecy throughput  is derived where cooperative jamming  is designed to confuse the relay    \\ [1ex]
 				\hline  
 				 \cite{lei2017secure, lei2018secrecy}  &MIMO-NOMA   &TAS-based security; secrecy outage probability/secrecy diversity order
 				over Nakagami-$m$ fading channels  are drived \\ [1ex]	
 				\hline		
 			\end{tabular}\label{table:pls}}
 	\end{center} 
 \end{table*}

%
%

 \subsection{AN-aided Secure NOMA}
 AN-aided approaches inject an artificial noise into directions orthogonal to those of the
main channel and are of great importance
when the eavesdropper's CSI is not available at the transmitter.
Different AN-aided secure NOMA systems are studied in \cite{lv2018secure,zhou2018artificial,zeng2019securing,liu2017enhancing,zheng2018secure}.
Specifically,  the secrecy outage performance of a multiple-input single-output (MISO) NOMA system is derived in \cite{lv2018secure}.
A MISO-NOMA-CR network using SWIPT is  studied in \cite{zhou2018artificial},
in which  AN-aided
cooperative jamming is used to improve
the security of the primary network.
Jamming signal is transmitted  by the cognitive BS
to cooperate with the primary BS to improve the security of the primary user.
Ergodic secrecy rates of the  users of an AN-aided massive MIMO-NOMA network is derived in \cite{zeng2019securing}. Asymptotic expressions  reveal that the AN
only affects the eavesdropper when the number of antennas is sufficiently large.

Using tools from stochastic geometry, a large-scale NOMA system is  studied in \cite{liu2017enhancing}
in which the BS can have single or multiple antennas while the users are assumed to have a single antenna.
For the multi-antenna BS, AN is generated at the BS. In both cases,  analytical expressions for the secrecy outage probability are derived.  Stochastic geometry based techniques are used
to model the locations of the legitimate users and eavesdropper, i.e., to deploy them
at spatially random locations.

 \subsection{Cooperative Transmission-based Secure NOMA}
Several independent works \cite{arafa2018secure,chen2018physical,xiang2019physical,zheng2018secure,yuan2019analysis} have  considered improving the security of NOMA-based
systems using cooperative transmission.
Analytical expressions for the secrecy
outage probability of a cooperative SISO-NOMA  with  an
eavesdropper and a single relay with both AF and DF protocols
are derived in \cite{chen2018physical}. In \cite{arafa2018trusted,arafa2018secure,arafa2018untrusted},  SISO-NOMA-based transmission
models with \textit{trusted} and \textit{untrusted}
cooperative relays are studied under cooperative jamming, AF, and DF  protocols.
In the case of trusted relays \cite{arafa2018trusted}, there are multiple cooperative relays, two legitimate users, and an external eavesdropper
whereas the number of relays in  the untrusted case \cite{arafa2018untrusted} is just one.
Achievable secrecy rate regions are derived in the above settings
show that the best relaying scheme  depends on different
parameters, such as the distances between the nodes and also on which part of the secrecy rate region
the system is to operate at.
  In particular, cooperative jamming is shown to outperform AF and DF  when the relays get closer
to the eavesdropper since the   jamming effect is more powerful in such a case.
Remarkably, in all the relaying schemes,  strictly positive secrecy rates are achievable
while secrecy rates can be zero without relays.
Secrecy in the context of NOMA VLC has been investigated in \cite{arafa2019vlc}, in which a two-user downlink is considered with an external eavesdropper. Under amplitude (peak power) constraints in the transmitted signals, imposed by the LEDs, achievable secrecy rate regions are derived with uniform signaling. Various relaying schemes are then shown to enhance the rate regions, also under peak power constraints at the relay luminaries.

Cooperative jamming is a popular and effective method  in improving the security in the PHY layer. In \cite{zheng2018secure}, a NOMA-based two-way relay network
 in the presence of single and multiple
eavesdroppers is studies. To ensure secure communications, the relay both forwards confidential information to the legitimate users  and  emits jamming signals.
An overlay NOMA-CR network with multiple primary and secondary users
under the secrecy constraint on primary users is proposed in \cite{xiang2019physical}  to improve the  secrecy outage performance.
Similar to \cite{zheng2018secure},  the motivation is to  give
the secondary transmitter the opportunity to access the spectrum
of primary users in exchange of relaying the message of them.
A cooperative relaying  scheme to enhancing the security of a  SISO-NOMA system  is considered in \cite{yuan2019analysis}. The source sends jamming signals while the relay is forwarding the message. In this model, there is no
external eavesdropper, but one of the NOMA users is seen as an eavesdropper. Cooperative jamming is also applied in \cite{arafa2018secure}, as explained previously.

 \subsection{Transmit Antenna Selection (TAS)-based Strategies}

MIMO systems are conceived to
increase reliability (on account of diversity) or to achieve high
data rates   (due to the multiplexing gain).
This performance  improvement of MIMO systems  can potentially scale up
with the number of antennas \cite{shin2016number}. However, the improvements  obtained
through MIMO  come at the price of a complex front-end architecture
and expensive RF chains.
Transmit antenna selection (TAS) is an effective technique that uses
a single RF chain (rather than multiple parallel RF chains) to
reduce cost, complexity, size, and power consumption while keeping the diversity and
throughput benefits acceptably high. This technique can also  improve security
in the presence of eavesdroppers  \cite{ferdinand2013effects,yang2013transmit,alves2012performance}.
 TAS has recently been applied  to improve the security of  MIMO-NOMA
systems in  \cite{lei2017secure, lei2018secrecy}.
Various TAS strategies are studied in \cite{lei2017secure} and
closed-form expressions for the secrecy outage probability are obtained.
In \cite{lei2018secrecy},  a max-min  TAS
strategy is utilized at the BS to improve the secure performance, and to obtain
the secrecy outage probability and to derive the available secrecy diversity order
over Nakagami-$m$ fading channels.
The literature of NOMA with PHY security is summarized in Table.~\ref{table:pls}.

 \subsection{Beamforming-based Strategies}

Beamforming is one of the most popular approaches in PHY security for multi-antenna systems.
ZF beamforming, which  is popular for the  multi-user MIMO systems \cite{wang2013slnr}, tries  to send the signal as orthogonal to the eavesdropper's channel
as possible.  When the transmitter has a larger number of antennas than the eavesdropper, it is possible to transmit the information
in the \textit{null space} of the eavesdropper's channel. This simple suboptimal approach tends to be asymptotically optimal in the massive MIMO case \cite{bjornson2016massive} and eliminates all the
inter-user interference under perfect CSI.
 Linear beamforming based on  the  \textit{generalized singular value decomposition} and   \textit{generalized eigenvalue decomposition} \cite{fakoorian2012optimal,khisti2010secureMISOME}, rotation-based beamforming \cite{vaezi2017optimal,zhang2019rot}, and numerical solutions are used to form the transmit covariance matrix for the secrecy rate maximization in the MISO and MIMO wiretap channels.

 \subsection{Future  Research Directions} 	
Most of the above PHY security solutions  are limited to the study of  two-user NOMA systems with  limited nodes. Extending those solutions to large-scale networks with multiple users in each cluster is of a great importance. Moreover, the extension of beamforming-based approaches to NOMA-based systems is not trivial and can
be seen as a good avenue for future research.
 In the case of beamforming and artificial noise based secrecy a few solutions with imperfect CSI exist. Further,  perfect SIC is assumed in the majority of NOMA PHY security papers, which may lead to
overestimating the performance of the networks considered.
Future  research may address the above issues.

\section{NOMA with Energy Harvesting} \label{sec:eh}

In this section, we discuss how energy harvesting technologies, mainly through SWIPT and WPCN, affect different performance metrics of NOMA, such as energy efficiency, achievable rates and outage probabilities, as reported in \cite{liu_noma_swipt_1, sun_noma_swipt_24, do_noma_swipt_10, xu_noma_swipt_6, ye_noma_swipt_14, ashraf_noma_swipt_11, do2_noma_swipt_15, alsaba_noma_swipt_19, diamantoulaki_noma_wpcn_2, song_noma_wpcn_22, pei_noma_wpcn_20, zewde_noma_wpcn_25, yang_noma_wpcn_consmp_18, wu_noma_wpcn_cnsmp_4, han_noma_rly_swipt_3, yang_noma_rly_swipt_13, zhang_noma_rly_swipt_21, dung_noma_rly_swipt_17,do2018csi, zhang_noma_rly_swipt_26, ha_noma_rly_swipt_5, kader_noma_swipt_cr_23, wang_noma_swipt_cr_12, zhao_noma_wit_wpt_16, xu_noma_eh_shrng_9}.

\subsection{NOMA with SWIPT} \label{sec:SWIPT}
In \cite{liu_noma_swipt_1, sun_noma_swipt_24, do_noma_swipt_10, xu_noma_swipt_6, ye_noma_swipt_14, ashraf_noma_swipt_11, do2_noma_swipt_15, alsaba_noma_swipt_19}, SWIPT is used to incentivize stronger downlink NOMA users, i.e., users that are relatively closer to the BS and have better channel conditions, to forward data to weaker users,
using energy harvested from the transmitted signals from the BS. Other works \cite{diamantoulaki_noma_wpcn_2, song_noma_wpcn_22, pei_noma_wpcn_20, zewde_noma_wpcn_25} focus on WPCN for NOMA uplink communications, in which the BS first transfers an amount of energy wirelessly to a number of users that is then used to let them communicate back to the BS in the uplink using NOMA. Similar ideas are investigated in \cite{yang_noma_wpcn_consmp_18, wu_noma_wpcn_cnsmp_4}, yet with the additional consideration of energy used in circuitry and hardware operations. Different from \cite{liu_noma_swipt_1, sun_noma_swipt_24, do_noma_swipt_10, xu_noma_swipt_6, ye_noma_swipt_14, ashraf_noma_swipt_11, do2_noma_swipt_15, alsaba_noma_swipt_19}, references \cite{han_noma_rly_swipt_3, yang_noma_rly_swipt_13, zhang_noma_rly_swipt_21, dung_noma_rly_swipt_17,do2018csi, zhang_noma_rly_swipt_26, ha_noma_rly_swipt_5} consider dedicated relaying nodes that rely on SWIPT to forward data to multiple users using NOMA. The works in \cite{kader_noma_swipt_cr_23, wang_noma_swipt_cr_12} use SWIPT within a cognitive radio NOMA framework. Designing modulation schemes for a two-user NOMA downlink, with an additional energy harvesting user, is considered in \cite{zhao_noma_wit_wpt_16}. The concept of energy cooperation among NOMA BSs is discussed in \cite{xu_noma_eh_shrng_9}. In what follows, we discuss some of the above-mentioned works in more detail.

\begin{figure}[t]
\center
\includegraphics[scale=.95]{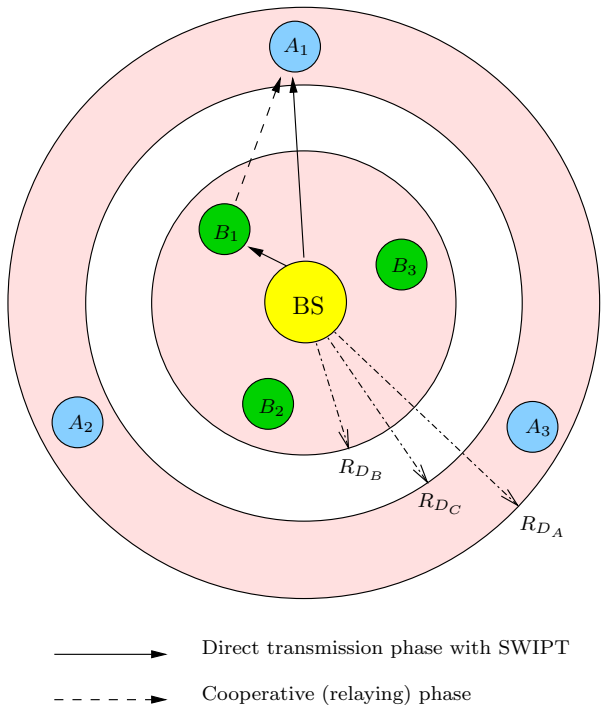}
\caption{A NOMA SWIPT system in which users close to the BS, labeled $\{B_i\}$, cooperatively assist the BS with transmitting data to further users, labeled $\{A_i\}$ \cite{liu_noma_swipt_1}.}
\label{fig:noma_swipt}
\end{figure}

Reference \cite{liu_noma_swipt_1} considers the model in Fig.~\ref{fig:noma_swipt}, in which users are categorized into near users, deployed within a disc of radius $R_{D_B}$ around the BS according to a homogeneous Poisson point process (PPP), and far users, deployed within a ring of inner and outer radii $R_{D_C}$ and $R_{D_A}$, respectively, according to another homogeneous PPP. It is assumed that $R_{D_C}\gg R_{D_B}$, and therefore near users assist the BS with relaying the far users' messages. Using SWIPT with power splitting, the near users harvest energy from the BS's signals in order to use it during the relaying phase. The far users then employ maximum ratio combining  to optimally combine the signals received from the BS and the near users.

Using this idea, reference \cite{liu_noma_swipt_1} presents some answers regarding how to pair near and far users together for such operation. Specifically, three user pairing strategies are investigated: 1) random near user and random far user (RNRF), in which pairing is assigned randomly; 2) nearest near user and nearest far user (NNNF), in which the nearest near and far users to the BS are paired; and 3) nearest near user and farthest far user (NNFF), in which a near user that is closest to the BS is paired with a far user that is farthest from the BS. It is shown that the best pairing strategy is NNNF, in the sense that it minimizes the outage probability and maximizes the achievable rates for both near and far users, which are derived in closed-form. This work concludes that by carefully choosing transmission rates and power splitting coefficients, one can achieve guaranteed performance results without the need to use the near users' own energy to power the relay phase transmission.

\begin{figure}[t]
\center
\includegraphics[scale=.95]{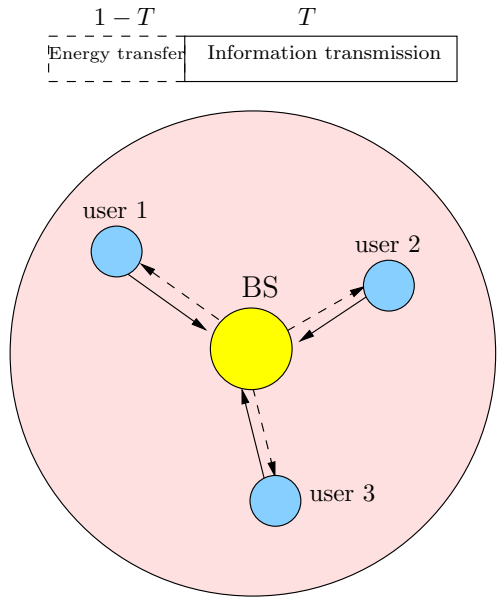}
\caption{Wireless energy transfer from the BS to the users to power the uplink transmission from the users to the BS via NOMA \cite{diamantoulaki_noma_wpcn_2}.}
\label{fig:noma_wpcn}
\end{figure}

\subsection{NOMA with WPCN} \label{sec:WPCN}

The model depicted in Fig.~\ref{fig:noma_wpcn} is considered in \cite{diamantoulaki_noma_wpcn_2}, in which a sequential energy transfer and information transmission is undertaken. The authors formulate different optimization problems to choose the best time duration $T$ that either maximizes the total achievable sum rate or maximizes the minimum achievable rate among the users so as to promote fairness. In addition, varying decoding orders are considered at the BS, where for a fraction $\tau_m$, $m=1,\dots,M$, of the time duration $T$ a specific decoding order is employed, with $M$ denoting the total number of possible decoding orders considered. This latter method is denoted by {\it time sharing}. The best time sharing vector ${\bm \tau}\triangleq[\tau_1,\dots,\tau_M]$ is then chosen to maximize either the sum or the minimum achievable rate, subject to $\sum_{m=1}^M\tau_m\leq1$. All considered problems are shown to be either linear or convex, which facilitates reaching an optimal solution in practice. One conclusion of this work is that implementing NOMA with WPCN offers an increase in user fairness, when compared to conventional OMA techniques, when the users' transmissions are constrained by the harvested energy.

\begin{figure*}[t]
	\center
	\includegraphics[scale=.9]{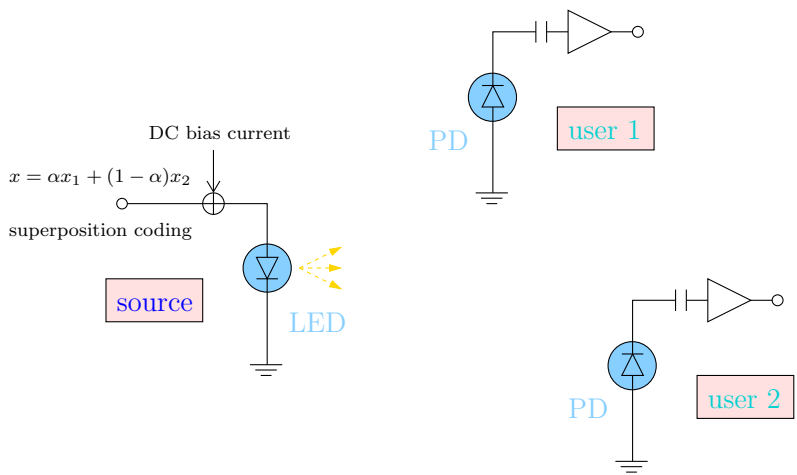}
	\caption{A two-user VLC NOMA downlink, in which superposition coding is used to drive the LED's DC bias current in order to send two messages to two users.}
	\label{fig_noma_vlc}
\end{figure*}

Reference \cite{wu_noma_wpcn_cnsmp_4} considers WPCN for uplink transmission, as in \cite{diamantoulaki_noma_wpcn_2}, yet with considering the energy consumed in circuitry operations. That is, it is assumed that each device consumes a constant amount of (passive) power, $p_c$, accounting for its transmit filter, mixer, synthesizer, etc. The authors compare two uplink transmission schemes following the wireless power transfer: TDMA (or OMA) and NOMA, in terms of energy and spectral efficiencies. It is shown that when $p_c>0$, TDMA is better in terms of both energy and spectral efficiency. It is therefore crucial to have an accurate power consumption model, especially for energy-constrained devices, before deciding on transmission schemes in general. In Table~\ref{table_eh}, we briefly mention the main technical contributions of the works that considered NOMA with energy harvesting in the recent literature.

\begin{table} [!t]
		\caption {Summary of technical contributions on NOMA with Energy Harvesting}
		\begin{center}{
				\begin{tabular}{|m{.6in}|m{2.2in}|@{}m{0cm}@{}}
					\hline 
					{\bf References} & {\bf Technical contribution} &\\ [1ex]
					\hline \hline
					\cite{liu_noma_swipt_1, sun_noma_swipt_24, do_noma_swipt_10, xu_noma_swipt_6, ye_noma_swipt_14, ashraf_noma_swipt_11, do2_noma_swipt_15, alsaba_noma_swipt_19}  & SWIPT to stronger NOMA users for relaying purposes to weaker users  &\\ [1ex]
					\hline
					\cite{diamantoulaki_noma_wpcn_2, song_noma_wpcn_22, pei_noma_wpcn_20, zewde_noma_wpcn_25} & WPCN to power NOMA users for uplink transmission  &\\ [1ex]
					\hline
					\cite{yang_noma_wpcn_consmp_18, wu_noma_wpcn_cnsmp_4} & Consideration of circuitry power consumptions  &\\ [1ex]
					\hline
					\cite{han_noma_rly_swipt_3, yang_noma_rly_swipt_13, zhang_noma_rly_swipt_21, dung_noma_rly_swipt_17, zhang_noma_rly_swipt_26, ha_noma_rly_swipt_5} & SWIPT to dedicated relaying nodes that forward data to NOMA users  &\\ [1ex]
					\hline
					\cite{kader_noma_swipt_cr_23, wang_noma_swipt_cr_12} & SWIPT in cognitive radio NOMA settings  &\\ [1ex]
					\hline
					\cite{zhao_noma_wit_wpt_16} & Modulation schemes design for additional energy harvesting users  &\\ [1ex]
					\hline
					\cite{ xu_noma_eh_shrng_9} & Energy cooperation among NOMA BSs  &\\ [1ex]
					\hline
					
				\end{tabular}\label{table_eh}}
	\end{center}
\end{table}

	\subsection{Future Research Directions}
	One interesting future direction in this line of research is to extend the concepts above to MIMO settings while considering the energy used in circuitry operations. While, in general, having multiple antennas delivers higher performance guarantees, it also consumes more circuitry power. Therefore, a tradeoff may exist in this situation between the number of antennas and the (passive) circuitry costs that needs to be optimally characterized. Another direction is to consider a combined SWIPT-WPCN settings, in which users divide their received energy from the BS into two main portions: one with which they relay data to other users in the network, and the other with which they communicate back to the BS in the uplink. Choosing the optimal portions in this situation would depend on the overall objective of the network. Finally, more practical signal design aspects for SWIPT should be taken into consideration as well. For instance, it is discussed in \cite{clerckx-swipt-nonlinear} that there can exist some dominant non-linear terms, in the transmitted signal power, that govern the amounts of received energy. This completely changes the transmitted signal design, and hence the communication rates as well. It would therefore be of interest to apply the ideas of \cite{clerckx-swipt-nonlinear} in the context of NOMA. Interference harvesting \cite{shin2017cooperative} is another appealing idea that can be applied in this context.

\section{NOMA with Visible Light Communications} \label{sec:vlc}

In this section, we discuss how NOMA techniques can be applied in the context of VLC, as reported in  \cite{kizilirmak_noma_vlc_2, zhang_noma_vlc_1, marshoud_noma_vlc_4, yin_noma_vlc_3, chen_noma_vlc_5, mitra_noma_vlc_12, yang_noma_vlc_6, marshoud_noma_vlc_7, yapici_noma_vlc_9, ma_noma_vlc_11, lin_noma_vlc_10, li_noma_vlc_13}. A typical two-user VLC NOMA downlink is shown in Fig.~\ref{fig_noma_vlc}.

\subsection{Existing Literature of NOMA-VLC }
	
The setting in which some $\epsilon$ fraction of the interference is not canceled at the strong user after employing SIC is considered in \cite{kizilirmak_noma_vlc_2, zhang_noma_vlc_1}. Reference \cite{kizilirmak_noma_vlc_2} considers a two-user downlink and compares NOMA-based VLC transmission with conventional OFDMA. Superiority of the achievable rate region using NOMA is shown. In \cite{zhang_noma_vlc_1}, the authors make use of the relatively high precise positioning in VLC to group NOMA users based on their locations. Within the group, SIC is employed for users with relatively better channel conditions. Two problems are formulated to decide on the power allocations among the users: one for maximizing the sum rate, and another for maximizing the minimum rate, subject to minimum QoS requirements for the users. The problems are solved iteratively after casting the problems into convex forms. 

The work in \cite{marshoud_noma_vlc_4} proposes a gain ratio power allocation (GPRA) technique to set the power allocated to each NOMA user based on its channel quality. The motivation is to ensure fairness among the users with low decoding order that suffer large interference. The effect of tuning the transmission angles of the LEDs is also studied.

Comparing NOMA to OFDMA is also undertaken in \cite{yin_noma_vlc_3}, in which outage probability expressions are derived for guaranteed QoS provisioning for a NOMA downlink, as well as ergodic sum rate expressions for opportunistic best effort service provisioning. The impact of LED lighting characteristics is also shown in this work.

MIMO-NOMA for VLC is considered in {\color{blue}\cite{chen_noma_vlc_5, mitra_noma_vlc_12}.}  In \cite{chen_noma_vlc_5}, normalized gain difference power allocation (NGDPA) strategies are employed, and detection is carried on via zero forcing followed by SIC. Improvements in achievable sum rates under the NGDPA strategy with respect to a generalization of the GPRA strategy of \cite{marshoud_noma_vlc_4} to the MIMO case, is shown via simulations. As the name suggests, in NGPDA, the power allocated is proportional to the normalized difference of channel gains between users, as opposed to the ratio of channel gains in GPRA. In \cite{mitra_noma_vlc_12}, a Chebyshev precoder is proposed based on singular value decomposition of the MIMO channel gain matrix for each NOMA user, in order to improve the performance of nonlinear LED compensation.

Peak power constraints, imposed by the LEDs to avoid clipping distortion, are considered in \cite{yang_noma_vlc_6, marshoud_noma_vlc_7}. Reference \cite{yang_noma_vlc_6} studies sum rate maximization problems, subject to user fairness and peak power constraints. Under logarithmic fairness, the proposed non-convex problems are efficiently converted to convex ones. Optimal power control algorithms are then derived using a Lagrangian framework and is shown to outperform other power allocation strategies, such as GPRA of \cite{marshoud_noma_vlc_4}. In \cite{marshoud_noma_vlc_7}, closed-form expressions for the bit error rate under perfect CSI are derived. The effect of delayed and noisy (imperfect) CSI is also studied.

User mobility is considered in \cite{yapici_noma_vlc_9, ma_noma_vlc_11}. Reference \cite{yapici_noma_vlc_9} considers the setting in which users randomly change their locations according to some distribution, and derives outage probability and sum rate expressions based on various NOMA users scheduling criteria. Both static and mobile users are considered in \cite{ma_noma_vlc_11}, for which upper and lower bounds on achievable rates are derived. Transmit power minimization problems are then formulated and solved by semi-definite relaxation techniques.

On a more practical aspect, an experimental demonstration is presented in \cite{lin_noma_vlc_10} for a bidirectional NOMA-OFDMA VLC network, and symmetric modulation techniques are proposed \cite{li_noma_vlc_13} to mitigate SIC error propagation in a downlink VLC-NOMA system.

\begin{table} [!t]
		\caption {Summary of technical contributions on NOMA with Visible Light Communications}
		\begin{center}{
				\begin{tabular}{|m{.6in}|m{2.2in}|@{}m{0cm}@{}}
					\hline
					{\bf References} & {\bf Technical contribution} &\\ [1ex]
					\hline \hline
					\cite{kizilirmak_noma_vlc_2, yin_noma_vlc_3}  & Superiority of VLC NOMA to OFDMA &\\ [1ex]
					\hline
					\cite{zhang_noma_vlc_1} & Grouping of NOMA users based on VLC positioning &\\ [1ex]
					\hline
					\cite{marshoud_noma_vlc_4} & Power allocation for fairness considerations &\\ [1ex]
					\hline
					\cite{chen_noma_vlc_5, mitra_noma_vlc_12} & MIMO VLC NOMA settings  &\\ [1ex]
					\hline
					\cite{yang_noma_vlc_6, marshoud_noma_vlc_7} & Amplitude (peak power) constraints for VLC NOMA systems  &\\ [1ex]
					\hline
					\cite{yapici_noma_vlc_9, ma_noma_vlc_11} & User mobility aspects and effects on VLC NOMA systems  &\\ [1ex]
					\hline
					\cite{lin_noma_vlc_10, li_noma_vlc_13} & Experimental demonstrations and practical modulation schemes  &\\ [1ex]
					\hline
				\end{tabular}\label{table_vlc}}
	\end{center}
\end{table}

In Table~\ref{table_vlc}, we briefly mention the main technical contributions of the works that considered NOMA with VLC in the recent literature.

	\subsection{Future Research Directions}
	Future directions for this line of research include a thorough investigation of optimal signaling schemes for VLC in the context of NOMA. Specifically, one differentiating aspect between VLC and other means of communications is the extra amplitude (peak power) set of constraints on the transmitted signals, which are imposed to maintain operation within the LEDs' dynamic range and to avoid clipping distortion. In these scenarios, Gaussian signaling is not even feasible, let alone optimal. Some candidate signaling schemes include: uniform, truncated Gaussian, and discrete signalings. A careful comparison between the rates achieved by these various signaling schemes is hence important, especially in MIMO settings, to realize NOMA VLC in practical scenarios.

\begin{figure*}[ht]
	\centering
	\includegraphics[width=0.9\linewidth]{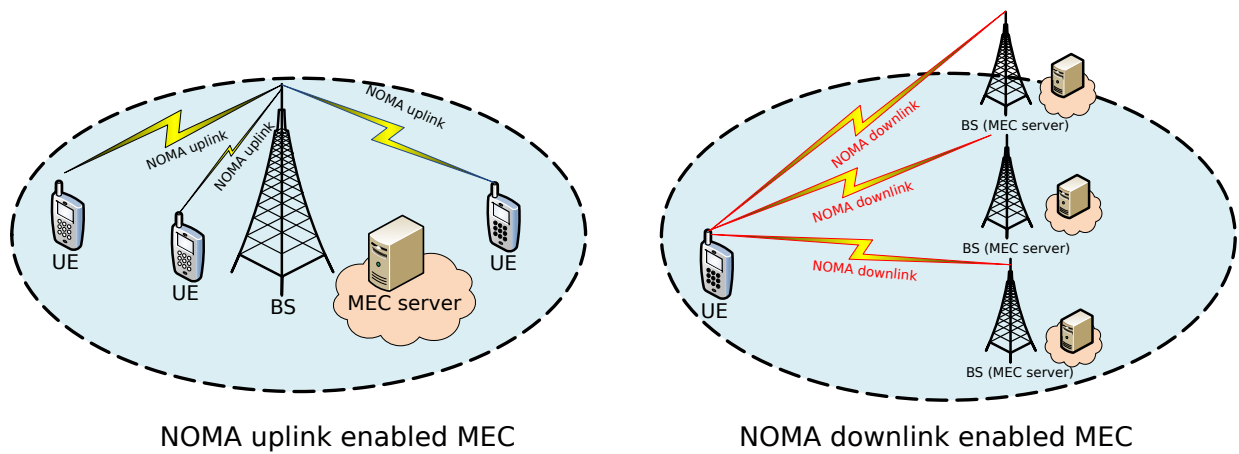}\\
	\caption{NOMA-MEC scenarios in the uplink and downlink.} \label{fig:mec}
\end{figure*}
\section{NOMA with Mobile Edge Computing}
\label{sec:mec}




Mobile edge computing (MEC)  brings computation/storage resources to  the mobile users at the edge of  network. As shown in Fig.\ref{fig:mec}, NOMA can be combined with  MEC both in the uplink and downlink. By applying  NOMA  into MEC, multiple users can offload their task simultaneously in the same frequency band. Therefore, the combination of NOMA and MEC can provide massive connectivity, low latency and high spectral efficiency.
\subsection{Combination of NOMA with MEC}
  Applying  uplink and downlink NOMA transmission into MEC offloading reveals that NOMA-MEC can provide lower latency and lower energy consumption than the traditional OMA scheme \cite{ZDing2018TCOM}. Resource optimization, however, plays an important role in NOMA-MEC, and as such, optimization of communication resource (e.g., offloading power, occupied frequency bandwidths, offloading time) and computing resource (e.g., task assignment and computing resource blocks) has recently attracted lots of  attention. Typical objectives of the resource optimization are energy consumption minimization \cite{FWang2018TCOM,ZDing2018WCL,AKiani2018JIOT} and task delay minimization \cite{ZDing2018WCLDely}. Among these works, \emph{partial} offloading and binary offloading are two main offloading schemes for NOMA-MEC. An energy efficient NOMA-MEC design is investigated in\cite{FWang2018TCOM}, in which both partial offloading and binary offloading are considered to minimize the weighted sum-energy consumption. Considering fully offloading scheme, where the task needs to be offloaded to the MEC server for remote computation, a hybrid NOMA-MEC system is proposed in \cite{ZDing2018WCL}. In this hybrid NOMA-MEC system, the user can offload parts of its task in the time-slot allocated to another user, and then offload the remaining task by its own allocated time-slot. The power allocation and offloading time are optimized to minimize the energy consumption  \cite{ZDing2018WCL}. Subsequently, the delay minimization is investigated on this hybrid NOMA-MEC system by using Dinkelbach method and Newton's method in \cite{ZDing2018WCLDely}, which reveals that these two methods converge to the same point, and Newton method converges faster than Dinkelbach method.  Further, closed-form expressions of optimal task assignment and transmit power were presented in \cite{FangTWC2019} to minimize the task delay in NOMA-MEC. In the binary offloading  NOMA-MEC,  communication  and  computing resource are jointly optimized to minimize the energy consumption for NOMA-MEC system \cite{AKiani2018JIOT}. When downlink NOMA transmission is applied into MEC systems, the mobile device can offload its tasks to several BSs equipped with MEC severs. By applying superposition coding at the mobile devices and SIC at BSs, the signals received at BSs has low interference. In this scenario, the transmit power and offloading task can be optimized to minimize the energy consumption of the NOMA enabled MEC system \cite{QGu2018Globecom}.

\subsection{Future Research Directions}
\label{SubSec:Potential_Works_NOMA}
 The following provides some potential research directions for NOMA-MEC.

\subsubsection{Joint computation and communication resource optimization} In the uplink transmission of NOMA-MEC, the BS equipped with MEC server has considerable computation resource to provide the mobile devices remote computation. Mobile devices can optimize the communication resource, i.e., transmit power and subcarrier allocation, to improve the performance of the system. Joint resource allocation is an inevitable trend to reduce the task delay and energy consumption of NOMA-MEC. In the downlink  transmission of NOMA-MEC,  computation capacity (i.e. central processing unit (CPU) frequency of MEC server or mobile device) and transmit power are also important factors to reduce the computation latency and energy consumption reduction.

\subsubsection{NOMA-MEC with imperfect CSI} Existing research in NOMA-MEC  mainly focus on  perfect CSI. Imperfect CSI is  one of the key obstacles in realizing the performance gain of NOMA in practice and may results in ambiguous decoding order of SIC. Improving the performance of NOMA-MEC with imperfect CSI including channel estimation errors, partial CSI, and limited channel feedback is an important research direction.
\subsubsection{Security in NOMA-MEC} When NOMA  is applied into MEC offloading,  secrecy and privacy  must be a concern since a passive or active eavesdropper may attempt to decode the mobile users' message. To address the scenario with external eavesdropper, PHY security can be utilized in NOMA-MEC system.
\subsubsection{Cooperative NOMA-MEC} When the mobile device is too far away from the main MEC server, the cooperative NOMA-MEC can be adapted to improve the connectivity of the network. In this scenario, one mobile device can help the far mobile device offload their tasks to the main MEC server. The mobile device transmits the superimposed signals to the primary MEC server and the helper, which acts as a relay helping MEC server. In this scenario, the performance analysis and resource allocation can be carried out for cooperative NOMA-MEC systems.

\begin{table*}[ht!]	
	\begin{center}{
	\caption{Summary of existing works on NOMA-MEC}
	\label{table:mec}
	\begin{tabular}{|p{1.9cm}|p{2.3cm}|p{.6 in}|p{9.5cm}|}
		\hline 
		\textbf{Topic} & \textbf{Frameworks} & \textbf{References}&\textbf{Contributions}\\ 
		\hline	\hline	\multirow{2}{2cm}{Architecture of
			\newline NOMA-MEC} & Uplink/downlink NOMA  MEC&\cite{ZDing2018TCOM}  & NOMA-MEC outperforms OMA MEC on lower latency and lower energy consumption than the traditional OMA scheme \\ 
		\hline
		\multirow{2}{2.2cm}{Energy consumption minimization} &Partial offloading&\cite{FWang2018TCOM,QGu2018Globecom}  & Resource allocation schemes are proposed to minimize the energy consumption for a uplink and downlink NOMA enabled MEC system \\ 
		\cline{2-4} 
		& Binary offloading
		& \cite{ZDing2018WCL, AKiani2018JIOT} & A hybrid NOMA-MEC system is proposed to minimize energy consumption \\
		\hline
		\multirow{1}{2.2cm}{\\Task delay minimization} &Partial offloading&\cite{FangTWC2019} & Optimal task and power allocation is proposed to minimize the task delay in NOMA-MEC system   \\ 
		\cline{2-4}
		& Binary offloading& \cite{ZDing2018WCLDely}& Dinkelbach and Newton’s methods are compared to minimize task delay for the hybrid NOMA-MEC system \\ 
		\hline		
	\end{tabular} }
\end{center} 
\end{table*}

	\section{NOMA with Other Technologies and Tools }
\label{sec:other}

Apart from the previously mentioned technologies, NOMA is also being  combined with many other technologies.   
Particularly, NOMA has been integrated to 
 vehicular communications  \cite{di2017v2x},  terrestrial-satellite communications \cite{vaezi2018book}, UAVs \cite{Liu2019WCM}, ambient backscatter communication \cite{zhang2019backscatter},
 wireless caching \cite{ding2018noma,doan2018optimal,xiang2019cache},  Wi-Fi networks \cite{khorov2018noma}, and so on. Typical challenges are very similar to the ones discussed earlier, and include clustering, power allocation, SIC performance, etc.  
	
Besides its great flexibility in being combined with other technologies, NOMA has successfully adopted different tools such as stochastic geometry, machine learning, and deep learning (DL) to solve large-scale and NP-hard problems that appear in related optimization problems 
in the uplink and downlink  NOMA \cite{cui2018unsupervised,zhang2018machine,xiao2018reinforcement,khai2018cash,luo2019deep}.
Particularly,  DL  has recently been applied to different problems. In \cite{khai2018cash},  DL is used for power allocation a caching based NOMA in three phases: exploration, training, and exploitation.
The exploration tries to learn which action returns the best reward
for each state, and gives a list of states and the corresponding best actions.  Due to its complexity, for the training phase,  deep neural networks are built.  Finally, the trained model is used to perform the power allocation for every state.
In \cite{luo2019deep}, DL is proposed to determine a solution of the joint downlink resource allocation problem for a SWIPT-enabled multi-carrier NOMA  system with a time switching-based receivers.

In \cite{khai2018cash},  DL is used for power allocation a caching based NOMA in three phases: exploration, training and exploitation.
 The exploration  tries to learn which action returns the best reward
 for each  state, and gives a list of states and the corresponding best actions.  Due to its complexity, for the training phase,  deep neural networks are built.  Finally, the trained model is used to perform the power allocation for every  state.
  In \cite{luo2019deep}, DL is proposed to determine a solution of the joint downlink resource allocation problem for a SWIPT-enabled multi carrier NOMA  system with a time switching-based receivers.

%
%
%
%
%
%
%
%
%
%
%
%
%
%
%
%
%
%
%
%
%
%
%

\section{Conclusion}
\label{sec:conc}
In this  paper, we have reviewed the research contributions of  NOMA combined with  several other technologies for 5G wireless networks and beyond, 
including  sub-6\,GHz and mmWave  massive MIMO, cognitive and cooperative communications, physical layer security, visible light
communications, energy harvesting, mobile edge computing, and machine learning and deep learning. It is elaborated how the combination of NOMA with these technologies can  overcome certain limits that those technologies are not able to overcome single-handedly. Challenges and future research directions in  line with this have also been discussed.


\begin{thebibliography}{100}
	\bibitem{vaezi2018book}
	M.~Vaezi, Z.~Ding, and H.~V. Poor, \emph{Multiple Access Techniques for 5G
		Wireless Networks and Beyond}.\hskip 1em plus 0.5em minus 0.4em\relax
	Springer, 2019.
	
	\bibitem{saito2013non}
	Y.~Saito, Y.~Kishiyama, A.~Benjebbour, T.~Nakamura, A.~Li, and K.~Higuchi,
	``{Non-orthogonal multiple access (NOMA) for cellular future radio access},''
	in \emph{Proc. IEEE VTC Spring}, 2013, pp. 1--5.
	
	\bibitem{ding2017application}
	Z.~Ding, Y.~Liu, J.~Choi, Q.~Sun, M.~Elkashlan, I.~Chih-Lin, and H.~V. Poor,
	``Application of non-orthogonal multiple access in {LTE} and {5G} networks,''
	\emph{{IEEE} Commun. Mag.}, vol.~55, no.~2, pp. 185--191, 2017.
	
	\bibitem{Cover}
	T.~M. Cover and J.~A. Thomas, \emph{{Elements of Information Theory}}.\hskip
	1em plus 0.5em minus 0.4em\relax New York: John Wiley \& Sons, 2006.
	
	\bibitem{ElGamal2011network}
	A.~El~Gamal and Y.~H. Kim, \emph{{Network Information Theory}}.\hskip 1em plus
	0.5em minus 0.4em\relax Cambridge University Press, 2011.
	
	\bibitem{tse2005fundamentals}
	D.~Tse and P.~Viswanath, \emph{Fundamentals of Wireless Communication}.\hskip
	1em plus 0.5em minus 0.4em\relax Cambridge university press, 2005.
	
	\bibitem{HK}
	T.~Han and K.~Kobayashi, ``{A new achievable rate region for the interference
		channel},'' \emph{{IEEE} Trans. Inf. Theory}, vol.~27, no.~1, pp. 49--60,
	January 1981.
	
	\bibitem{vaezi2016simplified}
	M.~Vaezi and H.~V. Poor, ``Simplified {Han-Kobayashi} region for one-sided and
	mixed {Gaussian} interference channels,'' in \emph{Proc. IEEE ICC}, 2016, pp.
	1--6.
	
	\bibitem{shin2017non}
	W.~Shin, M.~Vaezi, B.~Lee, D.~J. Love, J.~Lee, and H.~V. Poor,
	``{Non-orthogonal multiple access in multi-cell networks: Theory,
		performance, and practical challenges},'' \emph{{IEEE} Commun. Mag.},
	vol.~55, no.~10, pp. 176--183, 2017.
	
	\bibitem{vaezi2018non}
	M.~Vaezi, R.~Schober, Z.~Ding, and H.~V. Poor, ``Non-orthogonal multiple
	access: Common myths and critical questions,'' \emph{{IEEE} Wireless
		Commun.}, 2019.
	
	\bibitem{NOMA3GPP}
	{3GPP TD RP-150496}, ``{Study on Downlink Multiuser Superposition
		Transmission},'' Mar. 2015.
	
	\bibitem{Marzetta2010}
	T.~L. Marzetta, ``Noncooperative cellular wireless with unlimited numbers of
	base station antennas,'' \emph{{IEEE} Trans. Wireless Commun.}, vol.~9,
	no.~11, pp. 3590--3600, Nov. 2010.
	
	\bibitem{Lu2014}
	L.~Lu, G.~Y. Li, A.~L. Swindlehurst, A.~Ashikhmin, and R.~Zhang, ``An overview
	of massive {MIMO}: Benefits and challenges,'' \emph{{IEEE} J. Sel. Topic
		Signal Processing.}, vol.~8, no.~5, pp. 742--758, Oct. 2014.
	
	\bibitem{Rusek2013}
	F.~Rusek, D.~Persson, B.~K. Lau, E.~G. Larsson, T.~L. Marzetta, O.~Edfors, and
	F.~Tufvesson, ``Scaling up {MIMO}: {O}pportunities and challenges with very
	large arrays,'' \emph{{IEEE} Signal Process. Mag.}, vol.~30, no.~1, pp.
	40--60, Jan. 2013.
	
	\bibitem{Larsson2014}
	E.~G. Larsson, O.~Edfors, F.~Tufvesson, and T.~L. Marzetta, ``Massive {MIMO}
	for next generation wireless systems,'' \emph{{IEEE} Commun. Mag.}, vol.~52,
	no.~2, pp. 186--195, Feb. 2014.
	
	\bibitem{bjornson2016massive}
	E.~Bj{\"o}rnson, E.~G. Larsson, and T.~L. Marzetta, ``{Massive MIMO: Ten myths
		and one critical question},'' \emph{{IEEE} Commun. Mag.}, vol.~54, no.~2, pp.
	114--123, 2016.
	
	\bibitem{Senel2017}
	K.~Senel, H.~V. Cheng, E.~Bj{\"o}rnson, and E.~G. Larsson, ``What role can
	{NOMA} play in massive {MIMO}?'' \emph{{IEEE} J. Sel. Topic Signal
		Processing.}, Sep. 2018, {IEEE} {X}plore {E}arly {A}ccess.
	
	\bibitem{Rappaport2014}
	T.~S. Rappaport, R.~W. {Heath Jr.}, R.~C. Daniels, and J.~Murdock,
	\emph{Millimeter {W}ave {W}ireless {C}ommunications}.\hskip 1em plus 0.5em
	minus 0.4em\relax Prentice-{H}all, {E}nglewood {C}liffs, {NJ}, {USA}, 2014.
	
	\bibitem{Heath2016}
	R.~W. Heath, N.~González-Prelcic, S.~Rangan, W.~Roh, and A.~M. Sayeed, ``An
	overview of signal processing techniques for millimeter wave {MIMO}
	systems,'' \emph{{IEEE} J. Sel. Areas Commun.}, vol.~10, no.~3, pp. 436--453,
	Apr. 2016.
	
	\bibitem{Sendonaris2003a}
	A.~Sendonaris, E.~Erkip, and B.~Aazhang, ``User cooperation diversity. part i.
	system description,'' \emph{{IEEE} Trans. Commun.}, vol.~51, no.~11, pp.
	1927--1938, Nov. 2003.
	
	\bibitem{Sendonaris2003b}
	------, ``User cooperation diversity. part {II}. implementation aspects and
	performance analysis,'' \emph{{IEEE} Trans. Commun.}, vol.~51, no.~11, pp.
	1939--1948, Nov. 2003.
	
	\bibitem{Haykin2005}
	S.~Haykin, ``Cognitive radio: Brain-empowered wireless communications,''
	\emph{{IEEE} J. Sel. Areas Commun.}, vol.~23, no.~2, pp. 201--220, Feb. 2005.
	
	\bibitem{mukherjee2014principles}
	A.~Mukherjee, S.~A.~A. Fakoorian, J.~Huang, and A.~L. Swindlehurst,
	``Principles of physical layer security in multiuser wireless networks: A
	survey,'' \emph{{IEEE} Commun. Surv. Tut.}, vol.~16, no.~3, pp. 1550--1573,
	2014.
	
	\bibitem{shiu2011physical}
	Y.-S. Shiu, S.~Y. Chang, H.-C. Wu, S.~C.-H. Huang, and H.-H. Chen, ``{Physical
		layer security in wireless networks: A tutorial},'' \emph{{IEEE} Wireless
		Commun.}, vol.~18, no.~2, 2011.
	
	\bibitem{chen2017survey}
	X.~Chen, D.~W.~K. Ng, W.~Gerstacker, and H.-H. Chen, ``A survey on
	multiple-antenna techniques for physical layer security,'' \emph{{IEEE}
		Commun. Surv. Tut.}, 2017.
	
	\bibitem{liu2017physical}
	Y.~Liu, H.-H. Chen, and L.~Wang, ``{Physical layer security for next generation
		wireless networks: Theories, technologies, and challenges},'' \emph{{IEEE}
		Commun. Surv. Tut.}, vol.~19, no.~1, pp. 347--376, 2017.
	
	\bibitem{Ulukus2015}
	S.~{Ulukus}, A.~{Yener}, E.~{Erkip}, O.~{Simeone}, M.~{Zorzi}, P.~{Grover}, and
	K.~{Huang}, ``Energy harvesting wireless communications: A review of recent
	advances,'' \emph{{IEEE} J. Sel. Areas Commun.}, vol.~33, no.~3, pp.
	360--381, Mar. 2015.
	
	\bibitem{Pathak2015}
	P.~H. {Pathak}, X.~{Feng}, P.~{Hu}, and P.~{Mohapatra}, ``Visible light
	communication, networking, and sensing: A survey, potential and challenges,''
	\emph{{IEEE} Commun. Surveys Tuts.}, vol.~17, no.~4, pp. 2047--2077,
	Fourth-quarter 2015.
	
	\bibitem{PMMECSurvey2017}
	P.~Mach and Z.~Becvar, ``Mobile edge computing: A survey on architecture and
	computation offloading,'' \emph{IEEE Commun. Surveys Tuts.}, vol.~19, no.~3,
	pp. 1628--1656, 2017.
	
	\bibitem{wei2016survey}
	Z.~Ewi, J.~Yuan, W.~K.~N. Derrick, E.~Maged, and Z.~Ding, ``{A survey of
		downlink non-orthogonal multiple access for 5G wireless communication
		networks},'' \emph{ZTE Commun.}, vol.~14, pp. 17--25, 2016.
	
	\bibitem{islam2016power}
	S.~R. Islam, N.~Avazov, O.~A. Dobre, and K.-S. Kwak, ``{Power-domain
		non-orthogonal multiple access (NOMA) in 5G systems: Potentials and
		challenges},'' \emph{{IEEE} Commun. Surv. Tut.}, vol.~19, no.~2, pp.
	721--742, 2016.
	
	\bibitem{ding2017survey}
	Z.~Ding, X.~Lei, G.~K. Karagiannidis, R.~Schober, J.~Yuan, and V.~K. Bhargava,
	``{A survey on non-orthogonal multiple access for 5G networks: Research
		challenges and future trends},'' \emph{{IEEE} J. Sel. Areas Commun.},
	vol.~35, no.~10, pp. 2181--2195, 2017.
	
	\bibitem{Liu2017Proceeding}
	Y.~{Liu}, Z.~{Qin}, M.~{Elkashlan}, Z.~{Ding}, A.~{Nallanathan}, and
	L.~{Hanzo}, ``Nonorthogonal multiple access for {5G} and beyond,''
	\emph{Proc. IEEE}, vol. 105, no.~12, pp. 2347--2381, Dec. 2017.
	
	\bibitem{Sprint}
	Sprint unveils six {5G}-ready cities; significant milestone toward launching
	first {5G} mobile network in the {U.S.} [Online]. Available:
	\url{https://newsroom.sprint.com/sprint-unveils-5G-ready-massive-MIMO-markets.htm.}
	
	
	\bibitem{Chen2018a}
	X.~Chen, Z.~Zhang, C.~Zhong, R.~Jia, and D.~W.~K. Ng, ``Fully non-orthogonal
	communication for massive access,'' \emph{{IEEE} Trans. Commun.}, vol.~66,
	no.~4, pp. 1717--1731, Apr. 2018.
	
	\bibitem{Cheng2018b}
	H.~V. Cheng, E.~Bj{\"o}rnson, and E.~G. Larsson, ``Performance analysis of
	{NOMA} in training-based multiuser {MIMO} systems,'' \emph{{IEEE} Trans.
		Wireless Commun.}, vol.~17, no.~1, pp. 372--385, Jan. 2018.
	
	\bibitem{Kudathanthirige2019}
	D.~P. Kudathanthirige and G.~{Amarasuriya Aruma Baduge}, ``{NOMA}-aided
	multi-cell downlink massive {MIMO},'' \emph{{IEEE} J. Sel. Topic Signal
		Processing.}, Feb. 2019.
	
	\bibitem{Chen2018c}
	X.~Chen, F.~Gong, G.~Li, H.~Zhang, and P.~Song, ``User pairing and pair
	scheduling in massive {MIMO-NOMA} systems,'' \emph{{IEEE} Commun. Lett.},
	vol.~22, no.~4, pp. 788--791, Apr. 2018.
	
	\bibitem{Xu2017}
	C.~Xu, Y.~Hu, C.~Liang, J.~Ma, and L.~Ping, ``Massive {MIMO}, non-orthogonal
	multiple access and interleave division multiple access,'' \emph{{IEEE}
		{A}ccess}, vol.~5, pp. 14\,728--14\,748, Aug. 2017.
	
	\bibitem{Liu2018}
	L.~{Liu}, C.~{Yuen}, Y.~L. {Guan}, Y.~{Li}, and C.~{Huang}, ``Gaussian message
	passing for overloaded massive {MIMO-NOMA},'' \emph{{IEEE} Trans. Wireless
		Commun.}, vol.~18, no.~1, pp. 210--226, Jan 2019.
	
	\bibitem{LiuX2017}
	X.~Liu, Y.~Liu, X.~Wang, and H.~Lin, ``Highly efficient {3-D} resource
	allocation techniques in {5G} for {NOMA}-enabled massive {MIMO} and relaying
	systems,'' \emph{{IEEE} J. Sel. Areas Commun.}, vol.~35, no.~12, pp.
	2785--2797, Dec. 2017.
	
	\bibitem{Ma2017}
	J.~Ma, C.~Liang, C.~Xu, and L.~Ping, ``On orthogonal and superimposed pilot
	schemes in massive {MIMO NOMA} systems,'' \emph{{IEEE} J. Sel. Areas
		Commun.}, vol.~35, no.~12, pp. 2696--2707, Dec. 2017.
	
	\bibitem{Li2018b}
	Y.~{Li} and G.~{Amarasuriya Aruma Baduge}, ``{NOMA}-aided cell-free massive
	{MIMO} systems,'' \emph{IEEE Wireless Commun. Lett.}, vol.~7, no.~6, pp.
	950--953, Dec 2018.
	
	\bibitem{Li2019b}
	Y.~Li and G.~{Amarasuriya Aruma Baduge}, ``{NOMA}-aided massive {MIMO} downlink
	with distributed antenna arrays,'' in \emph{Proc. IEEE ICC}, May 2019, pp.
	1--7.
	
	\bibitem{Kusaladharma2019a}
	S.~Kusaladharma, W.~P. Zhu, W.~Ajib, and G.~{Amarasuriya Aruma Baduge},
	``Achievable rate analysis of {NOMA} in cell-free massive {MIMO}: A
	stochastic geometry approach,'' in \emph{Proc. IEEE ICC}, May 2019, pp. 1--7.
	
	\bibitem{Sharath2019a}
	S.~C.~R. Gaddam, D.~Kudathanthirige, and G.~{Amarasuriya Aruma Baduge},
	``Achievable rate analysis for {NOMA}-aided massive {MIMO} uplink,'' in
	\emph{Proc. IEEE ICC}, May 2019, pp. 1--7.
	
	\bibitem{Silva2019}
	S.~Silva, G.~{Amarasuriya Aruma Baduge}, C.~Tellambura, and M.~Ardakani,
	``{NOMA}-aided multi-way massive {MIMO} relay networks,'' in \emph{Proc. IEEE
		ICC}, May 2019, pp. 1--7.
	
	\bibitem{Aravinda2019}
	D.~Aravinda and G.~{Amarasuriya Aruma Baduge}, ``Cell-free massive {MIMO} with
	underlay spectrum-sharing,'' in \emph{Proc. IEEE ICC}, May 2019, pp. 1--7.
	
	\bibitem{Li2018a}
	Y.~{Li} and G.~{Amarasuriya Aruma Baduge}, ``Underlay spectrum-sharing massive
	{MIMO} {NOMA},'' \emph{{IEEE} Commun. Lett.}, vol.~23, no.~1, pp. 116--119,
	Jan 2019.
	
	\bibitem{Zhang2017a}
	D.~{Zhang}, Y.~{Liu}, Z.~{Ding}, Z.~{Zhou}, A.~{Nallanathan}, and T.~{Sato},
	``Performance analysis of non-regenerative massive-{MIMO-NOMA} relay systems
	for {5G},'' \emph{{IEEE} Trans. Commun.}, vol.~65, no.~11, pp. 4777--4790,
	Nov. 2017.
	
	\bibitem{Chen2018_Relay}
	X.~{Chen}, R.~{Jia}, and D.~W.~K. {Ng}, ``The application of relay to massive
	non-orthogonal multiple access,'' \emph{{IEEE} Trans. Commun.}, vol.~66,
	no.~11, pp. 5168--5180, Nov. 2018.
	
	\bibitem{Li2018c}
	Y.~Li and G.~{Amarasuriya Aruma Baduge}, ``Relay-aided massive {MIMO} {NOMA}
	downlink,'' in \emph{Proc. IEEE GLOBECOM}, Dec. 2018, pp. 1--7.
	
	\bibitem{Bjornson2018}
	E.~Bj{\"o}rnson, J.~Hoydis, and L.~Sanguinetti, ``Massive {MIMO} has unlimited
	capacity,'' \emph{{IEEE} Trans. Commun.}, vol.~17, no.~1, pp. 574--590, Jan.
	2018.
	
	\bibitem{Ho2018}
	C.~D. Ho, H.~Q. Ngo, M.~Matthaiou, and L.~D. Nguyen, ``Power allocation for
	multi-way massive {MIMO} relaying,'' \emph{{IEEE} Trans. Commun.}, vol.~66,
	no.~10, pp. 4457--4472, Oct. 2018.
	
	\bibitem{Rangan2014}
	S.~{Rangan}, T.~S. {Rappaport}, and E.~{Erkip}, ``Millimeter-wave cellular
	wireless networks: Potentials and challenges,'' \emph{Proc. IEEE}, vol. 102,
	no.~3, pp. 366--385, Mar. 2014.
	
	\bibitem{Rappaport2013}
	T.~S. {Rappaport}, S.~{Sun}, R.~{Mayzus}, H.~{Zhao}, Y.~{Azar}, K.~{Wang},
	G.~N. {Wong}, J.~K. {Schulz}, M.~{Samimi}, and F.~{Gutierrez}, ``Millimeter
	wave mobile communications for {5G} cellular: It will work{!}'' \emph{{IEEE}
		{A}ccess}, vol.~1, pp. 335--349, 2013.
	
	\bibitem{Molisch2015}
	X.~Zhang, A.~Molisch, and S.~Y. Kung, ``Variable-phase-shift-based
	{RF}-baseband codesign for {MIMO} antenna selection,'' \emph{IEEE Trans.
		Signal Process.}, vol.~53, no.~11, pp. 4091--4103, Nov. 2005.
	
	\bibitem{Sudarshan2006}
	P.~{Sudarshan}, N.~B. {Mehta}, A.~F. {Molisch}, and J.~{Zhang}, ``Channel
	statistics-based {RF} pre-processing with antenna selection,'' \emph{{IEEE}
		Trans. Wireless Commun.}, vol.~5, no.~12, pp. 3501--3511, Dec. 2006.
	
	\bibitem{Han2015}
	S.~{Han}, I.~{Chih-lin}, Z.~{Xu}, and C.~{Rowell}, ``Large-scale antenna
	systems with hybrid analog and digital beamforming for millimeter wave
	{5G},'' \emph{{IEEE} Commun. Mag.}, vol.~53, no.~1, pp. 186--194, Jan. 2015.
	
	\bibitem{Sohrabi2016}
	F.~{Sohrabi} and W.~{Yu}, ``Hybrid digital and analog beamforming design for
	large-scale antenna arrays,'' \emph{{IEEE} J. Sel. Topic Signal Processing.},
	vol.~10, no.~3, pp. 501--513, Apr. 2016.
	
	\bibitem{Wang2017}
	B.~Wang, L.~Dai, Z.~Wang, N.~Ge, and S.~Zhou, ``Spectrum and energy-efficient
	beamspace {MIMO-NOMA} for millimeter-wave communications using lens antenna
	array,'' \emph{{IEEE} J. Sel. Areas Commun.}, vol.~35, no.~10, pp.
	2370--2382, Oct. 2017.
	
	\bibitem{Wei2018}
	Z.~{Wei}, L.~{Zhao}, J.~{Guo}, D.~W.~K. {Ng}, and J.~{Yuan}, ``A multi-beam
	{NOMA} framework for hybrid mm{W}ave systems,'' in \emph{Proc. IEEE ICC}, May
	2018, pp. 1--7.
	
	\bibitem{Zhang2017_mmWave}
	D.~Zhang, Z.~Zhou, C.~Xu, Y.~Zhang, J.~Rodriguez, and T.~Sato, ``Capacity
	analysis of {NOMA} with mm{W}ave massive {MIMO} systems,'' \emph{{IEEE} J.
		Sel. Areas Commun.}, vol.~35, no.~7, pp. 1606--1618, Jul. 2017.
	
	\bibitem{Ding2017a}
	Z.~Ding and H.~V. Poor, ``Design of massive{-MIMO-NOMA} with limited
	feedback,'' \emph{{IEEE} Signal Process. Lett.}, vol.~23, no.~5, pp.
	629--633, May 2016.
	
	\bibitem{Ding2017b}
	Z.~Ding, L.~Dai, R.~Schober, and H.~V. Poor, ``{NOMA} meets finite resolution
	analog beamforming in massive {MIMO} and millimeter-wave networks,''
	\emph{{IEEE} Commun. Lett.}, vol.~21, no.~8, pp. 1879--1882, Aug. 2017.
	
	\bibitem{Mojtaba2018}
	M.~A. Almasi, M.~Vaezi, and H.~Mehrpouyan, ``Impact of beam misalignment on
	hybrid beamforming {NOMA} for {mmWave} communications,'' \emph{{IEEE} Trans.
		Commun.}, vol.~67, no.~6, pp. 4505--4518, Jun. 2019.
	
	\bibitem{almasi2019lens}
	M.~A. Almasi, R.~Amiri, M.~Vaezi, and H.~Mehrpouyan, ``Lens-based millimeter
	wave reconfigurable antenna {NOMA},'' in \emph{Proc. IEEE ICC Wkshps}, May
	2019, pp. 1--6.
	
	\bibitem{Wei2019}
	Z.~{Wei}, L.~{Zhao}, J.~{Guo}, D.~W.~K. {Ng}, and J.~{Yuan}, ``Multi-beam
	{NOMA} for hybrid mm{W}ave systems,'' \emph{{IEEE} Trans. Commun.}, vol.~67,
	no.~2, pp. 1705--1719, Feb. 2019.
	
	\bibitem{Wei2019a}
	Z.~{Wei}, D.~W. {Kwan Ng}, and J.~{Yuan}, ``{NOMA} for hybrid mm{W}ave
	communication systems with beamwidth control,'' \emph{{IEEE} J. Sel. Topic
		Signal Processing.}, vol.~13, no.~3, pp. 567--583, Jun. 2019.
	
	\bibitem{Brady2013}
	J.~Brady, N.~Behdad, and A.~Sayeed, ``Beamspace {MIMO} for millimeterwave
	communications: System architecture, modeling, analysis,and measurements,''
	\emph{IEEE Trans. Antennas Propag.}, vol.~61, no.~7, pp. 3814--3827, Jul.
	2013.
	
	\bibitem{Zhao2017}
	J.~{Zhao}, F.~{Gao}, W.~{Jia}, S.~{Zhang}, S.~{Jin}, and H.~{Lin}, ``Angle
	domain hybrid precoding and channel tracking for millimeter wave massive
	{MIMO} systems,'' \emph{{IEEE} Trans. Commun.}, vol.~16, no.~10, pp.
	6868--6880, Oct. 2017.
	
	\bibitem{Lin2017}
	H.~{Lin}, F.~{Gao}, S.~{Jin}, and G.~Y. {Li}, ``A new view of multi-user hybrid
	massive {MIMO}: {N}on-orthogonal angle division multiple access,''
	\emph{{IEEE} J. Sel. Areas Commun.}, vol.~35, no.~10, pp. 2268--2280, Oct.
	2017.
	
	\bibitem{Fan2017}
	D.~{Fan}, F.~{Gao}, G.~{Wang}, Z.~{Zhong}, and A.~{Nallanathan}, ``Angle domain
	signal processing-aided channel estimation for indoor {60-GHz} {TDD/FDD}
	massive {MIMO} systems,'' \emph{{IEEE} J. Sel. Areas Commun.}, vol.~35,
	no.~9, pp. 1948--1961, Sep. 2017.
	
	\bibitem{Van_Trees_2002}
	H.~L.~V. Trees, \emph{Optimum Array Processing: Part IV of Detection,
		Estimation, and Modulation Theory}.\hskip 1em plus 0.5em minus 0.4em\relax
	John Wiley and Sons, Inc., New York, 2002.
	
	\bibitem{ding2015cooperative}
	Z.~Ding, M.~Peng, and H.~V. Poor, ``{Cooperative non-orthogonal multiple access
		in 5{G} systems},'' \emph{{IEEE} Commun. Lett.}, vol.~19, no.~8, pp.
	1462--1465, 2015.
	
	\bibitem{Liu2016JSAC1}
	Y.~Liu, Z.~Ding, M.~Elkashlan, and H.~V. Poor, ``Cooperative non-orthogonal
	multiple access with simultaneous wireless information and power transfer,''
	\emph{{IEEE} J. Sel. Areas Commun.}, vol.~34, no.~4, pp. 938--953, Apr. 2016.
	
	\bibitem{Zhong2016CL}
	C.~{Zhong} and Z.~{Zhang}, ``Non-orthogonal multiple access with cooperative
	full-duplex relaying,'' \emph{{IEEE} Commun. Lett.}, vol.~20, no.~12, pp.
	2478--2481, Dec. 2016.
	
	\bibitem{Zhang2017TVT}
	Z.~{Zhang}, Z.~{Ma}, M.~{Xiao}, Z.~{Ding}, and P.~{Fan}, ``Full-duplex
	device-to-device-aided cooperative nonorthogonal multiple access,''
	\emph{{IEEE} Trans. Veh. Technol.}, vol.~66, no.~5, pp. 4467--4471, May 2017.
	
	\bibitem{Yue2018TCOM}
	X.~{Yue}, Y.~{Liu}, S.~{Kang}, A.~{Nallanathan}, and Z.~{Ding}, ``Exploiting
	full/half-duplex user relaying in {NOMA} systems,'' \emph{{IEEE} Trans.
		Commun.}, vol.~66, no.~2, pp. 560--575, Feb. 2018.
	
	\bibitem{Elbamby2017JSAC}
	M.~S. {Elbamby}, M.~{Bennis}, W.~{Saad}, M.~{Debbah}, and M.~{Latva-aho},
	``Resource optimization and power allocation in in-band full duplex-enabled
	non-orthogonal multiple access networks,'' \emph{{IEEE} J. Sel. Areas
		Commun.}, vol.~35, no.~12, pp. 2860--2873, Dec. 2017.
	
	\bibitem{Ding2016WCL}
	Z.~{Ding}, H.~{Dai}, and H.~V. {Poor}, ``Relay selection for cooperative
	{NOMA},'' \emph{{IEEE} Wireless Commun. Lett.}, vol.~5, no.~4, pp. 416--419,
	Aug. 2016.
	
	\bibitem{Men2015CL}
	J.~Men and J.~Ge, ``Non-orthogonal multiple access for multiple-antenna
	relaying networks,'' \emph{{IEEE} Commun. Lett.}, vol.~19, no.~10, pp.
	1686--1689, Oct. 2015.
	
	\bibitem{Shin2017SPL}
	W.~{Shin}, H.~{Yang}, M.~{Vaezi}, J.~{Lee}, and H.~V. {Poor}, ``Relay-aided
	{NOMA} in uplink cellular networks,'' \emph{{IEEE} Signal Process. Lett.},
	vol.~24, no.~12, pp. 1842--1846, Dec. 2017.
	
	\bibitem{Kim2015CoMP}
	J.~B. Kim and I.~H. Lee, ``Non-orthogonal multiple access in coordinated direct
	and relay transmission,'' \emph{{IEEE} Commun. Lett.}, vol.~19, no.~11, pp.
	2037--2040, Nov. 2015.
	
	\bibitem{Yue2018twoway}
	X.~{Yue}, Y.~{Liu}, S.~{Kang}, A.~{Nallanathan}, and Y.~{Chen}, ``Modeling and
	analysis of two-way relay non-orthogonal multiple access systems,''
	\emph{{IEEE} Trans. Commun.}, vol.~66, no.~9, pp. 3784--3796, Sep. 2018.
	
	\bibitem{Shin2017CL}
	W.~{Shin}, M.~{Vaezi}, B.~{Lee}, D.~J. {Love}, J.~{Lee}, and H.~V. {Poor},
	``Coordinated beamforming for multi-cell {MIMO-NOMA},'' \emph{{IEEE} Commun.
		Lett.}, vol.~21, no.~1, pp. 84--87, Jan. 2017.
	
	\bibitem{Liu2017JSAC}
	Y.~Liu, Z.~Qin, M.~Elkashlan, A.~Nallanathan, and J.~A. McCann,
	``Non-orthogonal multiple access in large-scale heterogeneous networks,''
	\emph{{IEEE} J. Sel. Areas Commun.}, vol.~35, no.~12, pp. 2667--2680, Dec.
	2017.
	
	\bibitem{Zhou2018Network}
	F.~{Zhou}, Y.~{Wu}, R.~Q. {Hu}, Y.~{Wang}, and K.~K. {Wong}, ``Energy-efficient
	{NOMA} enabled heterogeneous cloud radio access networks,'' \emph{IEEE
		Network}, vol.~32, no.~2, pp. 152--160, March 2018.
	
	\bibitem{Ali2018WCM}
	M.~S. {Ali}, E.~{Hossain}, and D.~I. {Kim}, ``Coordinated multipoint
	transmission in downlink multi-cell {NOMA} systems: Models and spectral
	efficiency performance,'' \emph{{IEEE} Wireless Commun.}, vol.~25, no.~2, pp.
	24--31, April 2018.
	
	\bibitem{Liu2018MIMONOMA}
	Y.~{Liu}, H.~{Xing}, C.~{Pan}, A.~{Nallanathan}, M.~{Elkashlan}, and
	L.~{Hanzo}, ``Multiple-antenna-assisted non-orthogonal multiple access,''
	\emph{{IEEE} Wireless Commun.}, vol.~25, no.~2, pp. 17--23, April 2018.
	
	\bibitem{shin2016number}
	W.~Shin, M.~Vaezi, J.~Lee, and H.~V. Poor, ``{On the number of users served in
		MIMO-NOMA cellular networks},'' in \emph{Proc. IEEE ISWCS}, 2016, pp.
	638--642.
	
	\bibitem{Gu2018CL}
	X.~{Gu}, X.~{Ji}, Z.~{Ding}, W.~{Wu}, and M.~{Peng}, ``Outage probability
	analysis of non-orthogonal multiple access in cloud radio access networks,''
	\emph{{IEEE} Commun. Lett.}, vol.~22, no.~1, pp. 149--152, Jan. 2018.
	
	\bibitem{Qin2018Mag}
	Z.~{Qin}, X.~{Yue}, Y.~{Liu}, Z.~{Ding}, and A.~{Nallanathan}, ``User
	association and resource allocation in unified {NOMA} enabled heterogeneous
	ultra dense networks,'' \emph{{IEEE} Commun. Mag.}, vol.~56, no.~6, pp.
	86--92, June 2018.
	
	\bibitem{Dai2015NOMA}
	L.~Dai, B.~Wang, Y.~Yuan, S.~Han, C.~l.~I, and Z.~Wang, ``Non-orthogonal
	multiple access for {5G}: solutions, challenges, opportunities, and future
	research trends,'' \emph{{IEEE} Commun. Mag.}, vol.~53, no.~9, pp. 74--81,
	Sep. 2015.
	
	\bibitem{jiang2017secure}
	M.~Jiang, Y.~Li, Q.~Zhang, Q.~Li, and J.~Qin, ``Secure beamforming in downlink
	{MIMO} nonorthogonal multiple access networks,'' \emph{{IEEE} Signal Process.
		Lett.}, vol.~24, no.~12, pp. 1852--1856, 2017.
	
	\bibitem{mitola2000cognitive}
	J.~Mitola, ``Cognitive radio---an integrated agent architecture for software
	defined radio,'' 2000.
	
	\bibitem{goldsmith09}
	A.~Goldsmith, S.~A. Jafar, I.~Maric, and S.~Srinivasa, ``Breaking spectrum
	gridlock with cognitive radios: An information theoretic perspective,''
	\emph{Proc. IEEE}, vol.~97, no.~5, pp. 894--914, May 2009.
	
	\bibitem{vaezi2012capacity}
	M.~Vaezi, ``The capacity of more capable cognitive interference channels,'' in
	\emph{Proc. IEEE Allerton}, 2014, pp. 372--377.
	
	\bibitem{Liu2015WPT}
	Y.~Liu, S.~A. Mousavifar, Y.~Deng, C.~Leung, and M.~Elkashlan, ``Wireless
	energy harvesting in a cognitive relay network,'' \emph{{IEEE} Trans.
		Wireless Commun.}, vol.~15, no.~4, pp. 2498--2508, Apr. 2016.
	
	\bibitem{QINTCOM2017}
	Z.~{Qin}, Y.~{Liu}, Y.~{Gao}, M.{Elkashlan}, and A.~{Nallanathan}, ``Wireless
	powered cognitive radio networks with compressive sensing and matrix
	completion,'' \emph{IEEE Trans. Commun.}, vol.~65, no.~4, pp. 1464--1476,
	Apr. 2017.
	
	\bibitem{QINSPM2018}
	Z.~{Qin}, J.~{Fan}, Y.~{Liu}, Y.~{Gao}, and G.~Y. {Li}, ``Sparse representation
	for wireless communications: A compressive sensing approach,'' \emph{IEEE
		Signal Process. Mag.}, vol.~35, no.~3, pp. 40--58, May 2018.
	
	\bibitem{Liu2016TVT}
	Y.~Liu, Z.~Ding, M.~Elkashlan, and J.~Yuan, ``Non-orthogonal multiple access in
	large-scale underlay cognitive radio networks,'' \emph{{IEEE} Trans. Veh.
		Technol.}, vol.~65, no.~12, pp. 10\,152--10\,157, Dec. 2016.
	
	\bibitem{Lv2017TCOM}
	L.~{Lv}, J.~{Chen}, Q.~{Ni}, and Z.~{Ding}, ``Design of cooperative
	non-orthogonal multicast cognitive multiple access for {5G} systems: User
	scheduling and performance analysis,'' \emph{IEEE Trans. Commun.}, vol.~65,
	no.~6, pp. 2641--2656, June 2017.
	
	\bibitem{Lv2018TVT}
	L.~{Lv}, L.~{Yang}, H.~{Jiang}, T.~H. {Luan}, and J.~{Chen}, ``When {NOMA}
	meets multiuser cognitive radio: Opportunistic cooperation and user
	scheduling,'' \emph{{IEEE} Trans. Veh. Technol.}, vol.~67, no.~7, pp.
	6679--6684, July 2018.
	
	\bibitem{ding2014pairing}
	Z.~Ding, P.~Fan, and H.~V. Poor, ``Impact of user pairing on {5G}
	non-orthogonal multiple access,'' \emph{{IEEE} Trans. Veh. Technol.},
	vol.~65, no.~8, pp. 6010--6023, Aug. 2016.
	
	\bibitem{Yang2016TWC}
	Z.~{Yang}, Z.~{Ding}, P.~{Fan}, and N.~{Al-Dhahir}, ``A general power
	allocation scheme to guarantee quality of service in downlink and uplink
	{NOMA} systems,'' \emph{{IEEE} Trans. Wireless Commun.}, vol.~15, no.~11, pp.
	7244--7257, Nov. 2016.
	
	\bibitem{Ding2016MIMO}
	Z.~Ding, L.~Dai, and H.~V. Poor, ``{MIMO-NOMA design for small packet
		transmission in the Internet of Things},'' \emph{IEEE Access}, vol.~4, pp.
	1393--1405, 2016.
	
	\bibitem{QIN:WM:2019}
	Z.~{Qin}, H.~{Ye}, G.~Y. {Li}, and B.~F. {Juang}, ``Deep learning in physical
	layer communications,'' \emph{IEEE Wireless Commun.}, vol.~26, no.~2, pp.
	93--99, Apr. 2019.
	
	\bibitem{liu2019machine}
	Y.~Liu, S.~Bi, Z.~Shi, and L.~Hanzo, ``When machine learning meets big data: A
	wireless communication perspective,'' \emph{arXiv:1901.08329}, 2019.
	
	\bibitem{QIN:IOT:2019}
	Z.~{Qin}, F.~Y. {Li}, G.~Y. {Li}, J.~{A. McCann}, and Q.~{Ni}, ``Low-power
	wide-area networks for sustainable {IoT},'' \emph{IEEE Wireless Commun.}, pp.
	1--6, 2019.
	
	\bibitem{Cui:JSTSP:2019}
	F.~{Cui}, Z.~{Qin}, Y.~{Cai}, M.~{Zhao}, and G.~Y. {Li}, ``Rethinking outage
	constraints for resource management in {NOMA} networks,'' \emph{{IEEE} J.
		Sel. Topic Signal Processing.}, vol.~13, no.~3, pp. 423--435, Jun. 2019.
	
	\bibitem{Liu2019WCM}
	Y.~{Liu}, Z.~{Qin}, Y.~{Cai}, Y.~{Gao}, G.~Y. {Li}, and A.~{Nallanathan},
	``{UAV} communications based on non-orthogonal multiple access,''
	\emph{{IEEE} Wireless Commun.}, vol.~26, no.~1, pp. 52--57, Feb. 2019.
	
	\bibitem{liang2019spectrum}
	L.~Liang, H.~Ye, and G.~Y. Li, ``Spectrum sharing in vehicular networks based
	on multi-agent reinforcement learning,'' \emph{arXiv preprint
		arXiv:1905.02910}, 2019.
	
	\bibitem{goel2008guaranteeing}
	S.~Goel and R.~Negi, ``Guaranteeing secrecy using artificial noise,''
	\emph{{IEEE} Trans. Wireless Commun.}, vol.~7, no.~6, 2008.
	
	\bibitem{fakoorian2012optimal}
	S.~A.~A. Fakoorian and A.~L. Swindlehurst, ``{Optimal power allocation for
		GSVD-based beamforming in the MIMO Gaussian wiretap channel},'' in
	\emph{Proc. IEEE ISIT}, 2012, pp. 2321--2325.
	
	\bibitem{shafiee2009towards}
	S.~Shafiee, N.~Liu, and S.~Ulukus, ``{Towards the secrecy capacity of the
		Gaussian MIMO wire-tap channel: The 2-2-1 channel},'' \emph{{IEEE} Trans.
		Inf. Theory}, vol.~55, no.~9, pp. 4033--4039, 2009.
	
	\bibitem{vaezi2017mimo}
	M.~Vaezi, W.~Shin, H.~V. Poor, and J.~Lee, ``{MIMO Gaussian wiretap channels
		with two transmit antennas: Optimal precoding and power allocation},'' in
	\emph{Proc. IEEE ISIT}, 2017, pp. 1708--1712.
	
	\bibitem{vaezi2017optimal}
	M.~Vaezi, W.~Shin, and H.~V. Poor, ``{Optimal beamforming for Gaussian MIMO
		wiretap channels with two transmit antennas},'' \emph{{IEEE} Trans. Wireless
		Commun.}, vol.~16, no.~10, pp. 6726--6735, October 2017.
	
	\bibitem{ferdinand2013effects}
	N.~S. Ferdinand, D.~B. da~Costa, and M.~Latva-aho, ``{Effects of outdated CSI
		on the secrecy performance of MISO wiretap channels with transmit antenna
		selection},'' \emph{{IEEE} Commun. Lett.}, vol.~17, no.~5, pp. 864--867,
	2013.
	
	\bibitem{yang2013transmit}
	N.~Yang, P.~L. Yeoh, M.~Elkashlan, R.~Schober, and I.~B. Collings, ``{Transmit
		antenna selection for security enhancement in MIMO wiretap channels},''
	\emph{{IEEE} Trans. Commun.}, vol.~61, no.~1, pp. 144--154, 2013.
	
	\bibitem{alves2012performance}
	H.~Alves, R.~D. Souza, M.~Debbah, and M.~Bennis, ``Performance of transmit
	antenna selection physical layer security schemes,'' \emph{{IEEE} Signal
		Process. Lett.}, vol.~19, no.~6, pp. 372--375, 2012.
	
	\bibitem{dong2010improving}
	L.~Dong, Z.~Han, A.~P. Petropulu, and H.~V. Poor, ``Improving wireless physical
	layer security via cooperating relays,'' \emph{{IEEE} Trans. Signal
		Process.}, vol.~58, no.~3, pp. 1875--1888, 2010.
	
	\bibitem{jeong2011optimal}
	C.~Jeong and I.-M. Kim, ``Optimal power allocation for secure multicarrier
	relay systems,'' \emph{{IEEE} Trans. Signal Process.}, vol.~59, no.~11, pp.
	5428--5442, 2011.
	
	\bibitem{li2011cooperative}
	J.~Li, A.~P. Petropulu, and S.~Weber, ``On cooperative relaying schemes for
	wireless physical layer security,'' \emph{{IEEE} Trans. Signal Process.},
	vol.~59, no.~10, pp. 4985--4997, 2011.
	
	\bibitem{vaezi2019noma}
	M.~Vaezi and H.~V. Poor, ``{NOMA: An Information-Theoretic Perspective},'' in
	\emph{Multiple Access Techniques for 5G Wireless Networks and Beyond}.\hskip
	1em plus 0.5em minus 0.4em\relax Springer, 2019, pp. 167--193.
	
	\bibitem{lv2018secure}
	L.~Lv, Z.~Ding, Q.~Ni, and J.~Chen, ``{Secure MISO-NOMA transmission with
		artificial noise},'' \emph{{IEEE} Trans. Veh. Technol.}, vol.~67, no.~7, pp.
	6700--6705, 2018.
	
	\bibitem{zhou2018artificial}
	F.~Zhou, Z.~Chu, H.~Sun, R.~Q. Hu, and L.~Hanzo, ``{Artificial noise aided
		secure cognitive beamforming for cooperative MISO-NOMA using SWIPT},''
	\emph{{IEEE} J. Sel. Areas Commun.}, vol.~36, no.~4, pp. 918--931, 2018.
	
	\bibitem{zeng2019securing}
	M.~Zeng, P.~Nguyen, O.~Dobre, and H.~V. Poor, ``{Securing downlink massive MIMO
		NOMA networks with artificial noise},'' \emph{{IEEE} J. Sel. Topic Signal
		Processing.}, 2019.
	
	\bibitem{liu2017enhancing}
	Y.~Liu, Z.~Qin, M.~Elkashlan, Y.~Gao, and L.~Hanzo, ``Enhancing the physical
	layer security of non-orthogonal multiple access in large-scale networks,''
	\emph{{IEEE} Trans. Wireless Commun.}, vol.~16, no.~3, pp. 1656--1672, 2017.
	
	\bibitem{zheng2018secure}
	B.~Zheng, M.~Wen, C.-X. Wang, X.~Wang, F.~Chen, J.~Tang, and F.~Ji, ``{Secure
		NOMA based two-way relay networks using artificial noise and full duplex},''
	\emph{{IEEE} J. Sel. Areas Commun.}, vol.~36, no.~7, pp. 1426--1440, 2018.
	
	\bibitem{arafa2018secure}
	A.~Arafa, W.~Shin, M.~Vaezi, and H.~V. Poor, ``Secure relaying in
	non-orthogonal multiple access: Trusted and untrusted scenarios,''
	\emph{arXiv preprint arXiv:1808.07864}, 2018.
	
	\bibitem{xiang2019physical}
	Z.~Xiang, W.~Yang, G.~Pan, Y.~Cai, and Y.~Song, ``{Physical layer security in
		cognitive radio inspired NOMA network},'' \emph{{IEEE} J. Sel. Topic Signal
		Processing.}, vol.~13, no.~3, pp. 700--714, 2019.
	
	\bibitem{yuan2019analysis}
	C.~Yuan, X.~Tao, N.~Li, W.~Ni, R.~P. Liu, and P.~Zhang, ``Analysis on secrecy
	capacity of cooperative non-orthogonal multiple access with proactive
	jamming,'' \emph{{IEEE} Trans. Veh. Technol.}, 2019.
	
	\bibitem{xiang2019secure}
	Z.~Xiang, W.~Yang, G.~Pan, Y.~Cai, and X.~Sun, ``Secure transmission in
	non-orthogonal multiple access networks with an untrusted relay,''
	\emph{{IEEE} Wireless Commun. Lett.}, 2019.
	
	\bibitem{lei2017secure}
	H.~Lei, J.~Zhang, K.-H. Park, P.~Xu, I.~S. Ansari, G.~Pan, B.~Alomair, and
	M.-S. Alouini, ``{On secure NOMA systems with transmit antenna selection
		schemes},'' \emph{IEEE Access}, vol.~5, pp. 17\,450--17\,464, 2017.
	
	\bibitem{lei2018secrecy}
	H.~Lei, J.~Zhang, K.-H. Park, P.~Xu, Z.~Zhang, G.~Pan, and M.-S. Alouini,
	``{Secrecy outage of Max--Min TAS scheme in MIMO-NOMA systems},''
	\emph{{IEEE} Trans. Veh. Technol.}, vol.~67, no.~8, pp. 6981--6990, 2018.
	
	\bibitem{chen2018physical}
	J.~Chen, L.~Yang, and M.-S. Alouini, ``{Physical layer security for cooperative
		NOMA systems},'' \emph{{IEEE} Trans. Veh. Technol.}, vol.~67, no.~5, pp.
	4645--4649, 2018.
	
	\bibitem{arafa2018trusted}
	A.~Arafa, W.~Shin, M.~Vaezi, and H.~V. Poor, ``{Securing downlink
		non-orthogonal multiple access systems by trusted relays},'' in \emph{Proc.
		IEEE GLOBECOM}, 2018, pp. 1--5.
	
	\bibitem{arafa2018untrusted}
	------, ``{Downlink non-orthogonal multiple access systems with an untrusted
		relay},'' in \emph{Proc. IEEE ACSSC}, 2018.
	
	\bibitem{arafa2019vlc}
	A.~Arafa, E.~Panayirci, and H.~V. Poor, ``Relay-aided secure broadcasting for
	visible light communications,'' \emph{{IEEE} Trans. Commun.}, vol.~67, 2019.
	
	\bibitem{wang2013slnr}
	K.~Wang, X.~Wang, and X.~Zhang, ``{SLNR-based transmit beamforming for MIMO
		wiretap channel},'' \emph{Wireless PersonalCcommunications}, vol.~71, no.~1,
	pp. 109--121, 2013.
	
	\bibitem{khisti2010secureMISOME}
	A.~Khisti and G.~W. Wornell, ``{Secure transmission with multiple antennas I:
		The MISOME wiretap channel},'' \emph{{IEEE} Trans. Inf. Theory}, vol.~56,
	no.~7, pp. 3088--3104, 2010.
	
	\bibitem{zhang2019rot}
	X.~Zhang, Y.~Qi, and M.~Vaezi, ``{A rotation-based method for precoding in
		Gaussian MIMOME channels},'' submitted for publication, 2019.
	
	\bibitem{liu_noma_swipt_1}
	Y.~Liu, Z.~Ding, M.~Elkashlan, and H.~V. Poor, ``Cooperative non-orthogonal
	multiple access with simultaneous wireless information and power transfer,''
	\emph{IEEE J. Sel. Areas Commun.}, vol.~34, no.~4, pp. 938--953, April 2016.
	
	\bibitem{sun_noma_swipt_24}
	R.~Sun, Y.~Wang, X.~Wang, and Y.~Zhang, ``Transceiver design for cooperative
	nonorthogonal multiple access systems with wireless energy transfer,''
	\emph{IET Commun.}, vol.~10, no.~15, pp. 1947--1955, October 2016.
	
	\bibitem{do_noma_swipt_10}
	N.~T. Do, D.~B. da~Costa, T.~Q. Duong, and B.~An, ``A {BNBF} user selection
	scheme for {NOMA}-based cooperative relaying systems with {SWIPT},''
	\emph{IEEE Commun. Lett.}, vol.~21, no.~3, pp. 664--667, March 2017.
	
	\bibitem{xu_noma_swipt_6}
	Y.~Xu, C.~Shen, Z.~Ding, X.~Sun, S.~Yan, and G.~Zhu, ``Joint beamforming design
	and power splitting control in cooperative {SWIPT} {NOMA} systems,'' in
	\emph{Proc. IEEE ICC}, May 2017.
	
	\bibitem{ye_noma_swipt_14}
	Y.~Ye, Y.~Li, D.~Wang, and G.~Lu, ``Power splitting protocol design for the
	cooperative {NOMA} with {SWIPT},'' in \emph{Proc. IEEE ICC}, May 2017.
	
	\bibitem{ashraf_noma_swipt_11}
	M.~Ashraf, A.~Shahid, J.~W. Jang, and K.-G. Lee, ``Energy harvesting
	non-orthogonal multiple access system with multi-antenna relay and base
	station,'' \emph{IEEE Access}, vol.~5, pp. 17\,660--17\,670, September 2017.
	
	\bibitem{do2_noma_swipt_15}
	T.~N. Do, D.~B. da~Costa, T.~Q. Duong, and B.~An, ``Improving the performance
	of cell-edge users in {MISO}-{NOMA} systems using {TAS} and {SWIPT}-based
	cooperative transmissions,'' \emph{IEEE Trans. Green Commun. Netw.}, vol.~2,
	no.~1, pp. 49--61, March 2018.
	
	\bibitem{alsaba_noma_swipt_19}
	Y.~Alsaba, C.~Y. Leow, and S.~K.~A. Rahim, ``Full-duplex cooperative
	non-orthogonal multiple access with beamforming and energy harvesting,''
	\emph{IEEE Access}, vol.~6, pp. 19\,726--19\,738, April 2018.
	
	\bibitem{diamantoulaki_noma_wpcn_2}
	P.~D. Diamantoulakis, K.~N. Pappi, Z.~Ding, and G.~K. Karagiannidis,
	``Wireless-powered communications with non-orthogonal multiple access,''
	\emph{IEEE Trans. Wireless Commun.}, vol.~15, no.~12, pp. 8422--8436,
	December 2016.
	
	\bibitem{song_noma_wpcn_22}
	M.~Song and M.~Zheng, ``Energy efficiency optimization for wireless powered
	sensor networks with nonorthogonal multiple access,'' \emph{IEEE Sens.
		Lett.}, vol.~2, no.~1, March 2018.
	
	\bibitem{pei_noma_wpcn_20}
	L.~Pei, Z.~Yang, C.~Pan, W.~Huang, M.~Chen, M.~Elkashlan, and A.~Nallanathan,
	``Energy-efficient {D2D} communications underlaying {NOMA}-based networks
	with energy harvesting,'' \emph{IEEE Commun. Lett.}, vol.~22, no.~5, pp.
	914--917, May 2018.
	
	\bibitem{zewde_noma_wpcn_25}
	T.~A. Zewde and M.~C. Gurosy, ``{NOMA}-based energy-efficient wireless powered
	communications,'' \emph{IEEE Trans. Green Commun. Netw.}, vol.~2, no.~3, pp.
	679--692, September 2018.
	
	\bibitem{yang_noma_wpcn_consmp_18}
	Z.~Yang, W.~Xu, Y.~Pan, C.~Pan, and M.~Chen, ``Energy efficient resource
	allocation in machine-to-machine communications with multiple access and
	energy harvesting for {IoT},'' \emph{IEEE Internet Things J.}, vol.~5, no.~1,
	pp. 229--245, February 2018.
	
	\bibitem{wu_noma_wpcn_cnsmp_4}
	Q.~Wu, W.~Chen, D.~W.~K. Ng, and R.~Schober, ``Spectral and energy-efficient
	wireless powered {IoT} networks: {NOMA} or {TDMA}?'' \emph{IEEE Trans. Veh.
		Technol.}, vol.~67, no.~7, pp. 6663--6667, July 2018.
	
	\bibitem{han_noma_rly_swipt_3}
	W.~Han, J.~Ge, and J.~Men, ``Performance analysis for {NOMA} energy harvesting
	relaying networks with transmit antenna selection and maximal-ratio combining
	over {N}akagami-m fading,'' \emph{IET Commun.}, vol.~10, no.~18, pp.
	2687--2693, December 2016.
	
	\bibitem{yang_noma_rly_swipt_13}
	Z.~Yang, Z.~Ding, P.~Fan, and N.~Al-Dhahir, ``The impact of power allocation on
	cooperative non-orthogonal multiple access networks with {SWIPT},''
	\emph{IEEE Trans. Wireless Commun.}, vol.~16, no.~7, pp. 4332--4343, July
	2017.
	
	\bibitem{zhang_noma_rly_swipt_21}
	Y.~Zhang and J.~Ge, ``Performance analysis for non-orthogonal multiple access
	in energy harvesting relaying networks,'' \emph{IET Commun.}, vol.~11,
	no.~11, pp. 1768--1774, August 2017.
	
	\bibitem{dung_noma_rly_swipt_17}
	L.~T. Dung, T.~M. Hoang, N.~T. Tan, and S.-G. Choi, ``Analysis of partial relay
	selection in {NOMA} systems with {RF} energy harvesting,'' in \emph{Proc.
		SigTelCom}, January 2018.
	
	\bibitem{do2018csi}
	D.~H. Do, M.~Vaezi, and T.~L. Nguyen, ``{Wireless powered cooperative relaying
		using NOMA with imperfect CSI},'' in \emph{Proc. IEEE GLOBECOM Wkshps}, 2018,
	pp. 1--5.
	
	\bibitem{zhang_noma_rly_swipt_26}
	Y.~Zhang and J.~Ge, ``Impact analysis for user pairing on {NOMA}-based energy
	harvesting relaying networks with imperfect {CSI},'' \emph{IET Commun.},
	vol.~12, no.~13, pp. 1609--1614, August 2018.
	
	\bibitem{ha_noma_rly_swipt_5}
	D.-B. Ha and S.~Q. Nguyen, ``Outage performance of energy harvesting {DF}
	relaying {NOMA} networks,'' \emph{Mobile Netw. Appl.}, vol.~23, no.~6, pp.
	1572--1585, December 2018.
	
	\bibitem{kader_noma_swipt_cr_23}
	M.~F. Kader, M.~B. Shahab, and S.~Y. Shin, ``Cooperative spectrum sharing with
	energy harvesting best secondary user selection and non-orthogonal multiple
	access,'' in \emph{Proc. IEEE ICNC}, January 2017.
	
	\bibitem{wang_noma_swipt_cr_12}
	Y.~Wang, Y.~Wu, F.~Zhou, Z.~Chu, Y.~Wu, and F.~Yuan, ``Multi-objective resource
	allocation in a {NOMA} cognitive radio network with a practical non-linear
	energy harvesting model,'' \emph{IEEE Access}, vol.~6, pp. 12\,973--12\,982,
	March 2018.
	
	\bibitem{zhao_noma_wit_wpt_16}
	Y.~Zhao, J.~Hu, Z.~Ding, and K.~Yang, ``Constellation rotation aided modulation
	design for the multi-user {SWIPT}-{NOMA},'' in \emph{Proc. IEEE ICC}, May
	2018.
	
	\bibitem{xu_noma_eh_shrng_9}
	B.~Xu, Y.~Chen, J.~R. Carri\'{o}n, and T.~Zhang, ``Resource allocation in
	energy-cooperation enabled two-tier {NOMA} {HetNets} toward green {5G},''
	\emph{IEEE J. Sel. Areas Commun.}, vol.~35, no.~12, pp. 2758--2770, December
	2017.
	
	\bibitem{clerckx-swipt-nonlinear}
	B.~Clerckx, R.~Zhang, R.~Schober, D.~W.~K. Ng, D.~I. Kim, and H.~V. Poor,
	``Fundamentals of wireless information and power transfer: From rf energy
	harvester models to signal and system designs,'' \emph{IEEE J. Sel. Areas
		Commun.}, vol.~37, no.~1, pp. 4--33, January 2019.
	
	\bibitem{shin2017cooperative}
	W.~Shin, M.~Vaezi, J.~Lee, and H.~V. Poor, ``Cooperative wireless powered
	communication networks with interference harvesting,'' \emph{{IEEE} Trans.
		Veh. Technol.}, vol.~67, no.~4, pp. 3701--3705, 2017.
	
	\bibitem{kizilirmak_noma_vlc_2}
	R.~C. Kizilirmak, C.~R. Rowell, and M.~Uysal, ``Non-orthogonal multiple access
	({NOMA}) for indoor visible light communications,'' in \emph{Proc. IWOW},
	September 2015.
	
	\bibitem{zhang_noma_vlc_1}
	X.~Zhang, Q.~Gao, C.~Gong, and Z.~Xu, ``User grouping and power allocation for
	{NOMA} visible light communication multi-cell networks,'' \emph{IEEE Commun.
		Lett.}, vol.~21, no.~4, pp. 777--780, April 2017.
	
	\bibitem{marshoud_noma_vlc_4}
	H.~Marshoud, V.~M. Kapinas, G.~K. Karagiannidis, and S.~Muhaidat,
	``Non-orthogonal multiple access for visible light communications,''
	\emph{IEEE Photon. Technol. Lett.}, vol.~28, no.~1, pp. 51--54, January 2016.
	
	\bibitem{yin_noma_vlc_3}
	L.~Yin, W.~O. Popoola, X.~Wu, and H.~Haas, ``Performance evaluation of
	non-orthogonal multiple access in visible light communication,'' \emph{IEEE
		Trans. Commun.}, vol.~64, no.~12, pp. 5162--5175, December 2016.
	
	\bibitem{chen_noma_vlc_5}
	C.~Chen, W.-D. Zhong, and H.~Yang, ``On the performance of {MIMO}-{NOMA}-based
	visible light communication systems,'' \emph{IEEE Photon. Technol. Lett.},
	vol.~30, no.~4, pp. 307--310, February 2018.
	
	\bibitem{mitra_noma_vlc_12}
	R.~Mitra and V.~Bhatia, ``Precoded {C}hebyshev-{NLMS}-based pre-distorter for
	nonlinear {LED} compensation in {NOMA}-{VLC},'' \emph{IEEE Trans. Commun.},
	vol.~65, no.~11, pp. 4845--4856, November 2017.
	
	\bibitem{yang_noma_vlc_6}
	Z.~Yang, W.~Xu, and Y.~Li, ``Fair non-orthogonal multiple access for visible
	light communication downlinks,'' \emph{IEEE Wireless Commun. Lett.}, vol.~6,
	no.~1, pp. 66--69, February 2017.
	
	\bibitem{marshoud_noma_vlc_7}
	H.~Marshoud, P.~C. Sofotasios, S.~Muhaidat, G.~K. Karagiannidis, and B.~S.
	Sharif, ``On the performance of visible light communication systems with
	non-orthogonal multiple access,'' \emph{IEEE Trans. Wireless Commun.},
	vol.~16, no.~10, pp. 6350--6364, October 2017.
	
	\bibitem{yapici_noma_vlc_9}
	Y.~Yapici and I.~Guvenc, ``Non-orthogonal multiple access for mobile {VLC}
	networks with random receiver orientation,'' arXiv1801.04888.
	
	\bibitem{ma_noma_vlc_11}
	S.~Ma, Y.~He, H.~Li, S.~Lu, F.~Zhang, and S.~Li, ``Optimal power allocation for
	mobile users in non-orthogonal multiple access visible light communication
	networks,'' \emph{IEEE Trans. Commun.}, to appear.
	
	\bibitem{lin_noma_vlc_10}
	B.~Lin, W.~Ye, X.~Tang, and Z.~Ghassemlooy, ``Experimental demonstration of
	bidirectional {NOMA}-{OFDMA} visible light communications,'' \emph{Oprics
		Express}, vol.~25, no.~4, pp. 4348--4355, February 2017.
	
	\bibitem{li_noma_vlc_13}
	H.~Li, Z.~Huang, Y.~Xiao, S.~Zhan, and Y.~Ji, ``Solution for error propagation
	in a {NOMA}-based {VLC} network: symmetric superposition coding,''
	\emph{Oprics Express}, vol.~25, no.~24, pp. 29\,856--29\,863, November 2017.
	
	\bibitem{ZDing2018TCOM}
	Z.~{Ding}, P.~{Fan}, and H.~V. {Poor}, ``Impact of non-orthogonal multiple
	access on the offloading of mobile edge computing,'' \emph{IEEE Trans.
		Commun.}, vol.~67, no.~1, pp. 375--390, Jan. 2019.
	
	\bibitem{FWang2018TCOM}
	F.~{Wang}, J.~{Xu}, and Z.~{Ding}, ``Multi-antenna {NOMA} for computation
	offloading in multiuser mobile edge computing systems,'' \emph{IEEE Trans.
		Commun.}, vol.~67, no.~3, pp. 2450--2463, Mar. 2019.
	
	\bibitem{ZDing2018WCL}
	Z.~Ding, J.~Xu, O.~A. Dobre, and H.~V. Poor, ``Joint power and time allocation
	for {NOMA-MEC} offloading,'' \emph{Submitted to IEEE Wireless. Commun.
		Lett.}, 2018.
	
	\bibitem{AKiani2018JIOT}
	A.~Kiani and N.~Ansari, ``Edge computing aware {NOMA} for 5{G} networks,''
	\emph{IEEE Internet Things}, vol.~5, no.~2, pp. 1299--1306, Apr. 2018.
	
	\bibitem{ZDing2018WCLDely}
	Z.~{Ding}, D.~W.~K. {Ng}, R.~{Schober}, and H.~V. {Poor}, ``Delay minimization
	for {NOMA-MEC} offloading,'' \emph{IEEE Signal Process. Lett.}, vol.~25,
	no.~12, pp. 1875--1879, Dec. 2018.
	
	\bibitem{FangTWC2019}
	F.~Fang, Y.~Xu, Z.~Ding, C.~Shen, M.~Peng, and G.~K. Karagiannidis, ``Optimal
	task assignment and power allocation for {NOMA} mobile-edge computing
	networks,'' arXiv:1904.12389.
	
	\bibitem{QGu2018Globecom}
	Q.~{Gu}, G.~{Wang}, J.~{Liu}, R.~{Fan}, D.~{Fan}, and Z.~{Zhong}, ``Optimal
	offloading with non-orthogonal multiple access in mobile edge computing,'' in
	\emph{Proc. IEEE GLOBECOM}, Dec. 2018, pp. 1--5.
	
	\bibitem{di2017v2x}
	B.~Di, L.~Song, Y.~Li, and Z.~Han, ``{V2X meets NOMA: Non-orthogonal multiple
		access for 5G-enabled vehicular networks},'' \emph{IEEE Wireless Commun.},
	vol.~24, no.~6, pp. 14--21, 2017.
	
	\bibitem{zhang2019backscatter}
	Q.~Zhang, L.~Zhang, Y.-C. Liang, and P.-Y. Kam, ``{Backscatter-NOMA: A
		symbiotic system of cellular and Internet-of-Things networks},'' \emph{IEEE
		Access}, vol.~7, pp. 20\,000--20\,013, 2019.
	
	\bibitem{ding2018noma}
	Z.~Ding, P.~Fan, G.~K. Karagiannidis, R.~Schober, and H.~V. Poor, ``{NOMA
		assisted wireless caching: Strategies and performance analysis},''
	\emph{{IEEE} Trans. Commun.}, vol.~66, no.~10, pp. 4854--4876, 2018.
	
	\bibitem{doan2018optimal}
	K.~N. Doan, W.~Shin, M.~Vaezi, H.~V. Poor, and T.~Q. Quek, ``Optimal power
	allocation in cache-aided non-orthogonal multiple access systems,'' in
	\emph{Proc. IEEE ICC Workshops}, 2018, pp. 1--6.
	
	\bibitem{xiang2019cache}
	L.~Xiang, D.~W.~K. Ng, X.~Ge, Z.~Ding, V.~W. Wong, and R.~Schober,
	``Cache-aided non-orthogonal multiple access: The two-user case,''
	\emph{{IEEE} J. Sel. Topic Signal Processing.}, vol.~13, no.~3, pp. 436--451,
	2019.
	
	\bibitem{khorov2018noma}
	E.~Khorov, A.~Kureev, and I.~Levitsky, ``{NOMA testbed on Wi-Fi},'' in
	\emph{Proc. IEEE PIMRC}, 2018, pp. 1153--1154.
	
	\bibitem{cui2018unsupervised}
	J.~Cui, Z.~Ding, P.~Fan, and N.~Al-Dhahir, ``Unsupervised machine
	learning-based user clustering in millimeter-wave-{NOMA} systems,''
	\emph{{IEEE} Trans. Wireless Commun.}, vol.~17, no.~11, pp. 7425--7440, 2018.
	
	\bibitem{zhang2018machine}
	N.~Zhang, K.~Cheng, and G.~Kang, ``A machine-learning-based blind detection on
	interference modulation order in {NOMA} systems,'' \emph{{IEEE} Commun.
		Lett.}, vol.~22, no.~12, pp. 2463--2466, 2018.
	
	\bibitem{xiao2018reinforcement}
	L.~Xiao, Y.~Li, C.~Dai, H.~Dai, and H.~V. Poor, ``Reinforcement learning-based
	{NOMA} power allocation in the presence of smart jamming,'' \emph{{IEEE}
		Trans. Veh. Technol.}, vol.~67, no.~4, pp. 3377--3389, 2018.
	
	\bibitem{khai2018cash}
	K.~D. Nguyen, M.~Vaezi, W.~Shin, H.~V. Poor, H.~Shin, and T.~Q. Quek, ``{Power
		allocations in cache-aided NOMA systems: Optimization and deep learning
		approaches},'' submitted for publication, 2019.
	
	\bibitem{luo2019deep}
	J.~Luo, J.~Tang, D.~So, G.~Chen, K.~Cumanan, and J.~Chambers, ``A deep
	learning-based approach to power minimization in multi-carrier {NOMA} with
	{SWIPT},'' \emph{IEEE Access}, pp. 17\,450--17\,460, 2019.
	
\end{thebibliography}

\begin{IEEEbiography}[{\includegraphics[width=1in,height=1.25in,clip,keepaspectratio]{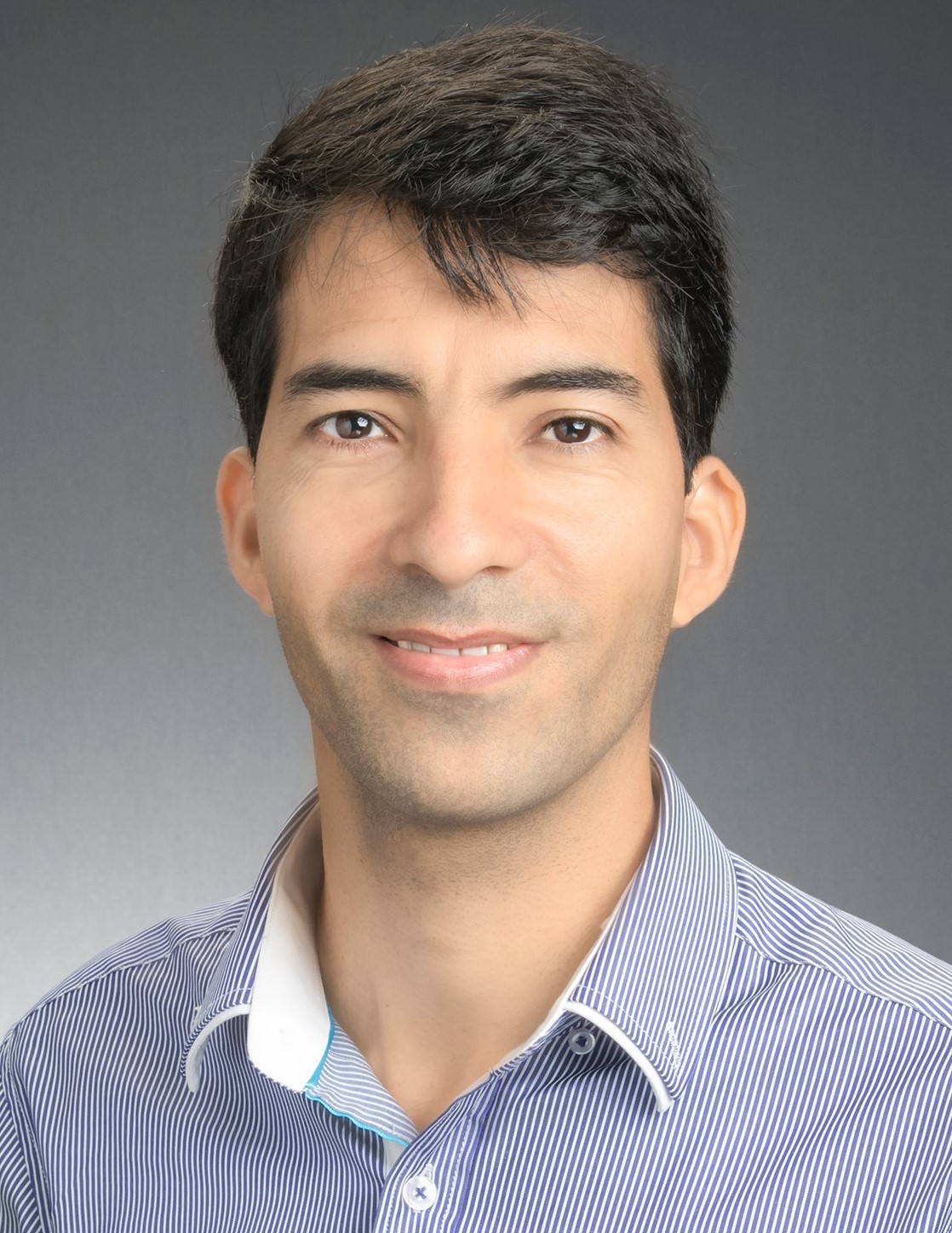}}]{Mojtaba Vaezi}  (S'09, M'14, SM'18)
	received the Ph.D. degree  from McGill University in 2014
	and the B.Sc. and M.Sc. degrees from Amirkabir University of Technology (Tehran Polytechnic), Iran, all in Electrical Engineering.
	From 2015-2018 he was a Postdoctoral Research Fellow and Associate Research Scholar at Princeton University. He is currently an Assistant Professor of ECE at Villanova University.
	Before joining Princeton, he was a researcher at Ericsson Research in Montreal, Canada. His research interests include the broad areas of   signal processing and machine learning for wireless communications   with an emphasis on physical layer security and fifth generation (5G) radio access technologies.   Among his publications in these areas is the book \textit{Multiple Access Techniques for 5G Wireless Networks and Beyond} (Springer, 2019).
	
	Dr. Vaezi is an Editor of \textit{IEEE Transactions on Communications} and \textit{IEEE Communications Letters}. He has co-organized five NOMA workshops at  IEEE VTC 2017-Spring,  Globecom'17, 18 and ICC'18, 19. He is a recipient of several academic, leadership, and research awards, including McGill Engineering Doctoral Award, IEEE Larry K. Wilson Regional Student Activities Award in 2013,  the Natural Sciences and Engineering Research Council of Canada (NSERC) Postdoctoral Fellowship in 2014,  Ministry of Science and ICT of Korea's  best paper award  in 2017, and IEEE Communications Letters Exemplary Editor Award in 2018.
\end{IEEEbiography}

\vspace{-1.1cm}

\begin{IEEEbiography}
	[{\includegraphics[width=1in,height=1.25in,clip,keepaspectratio]{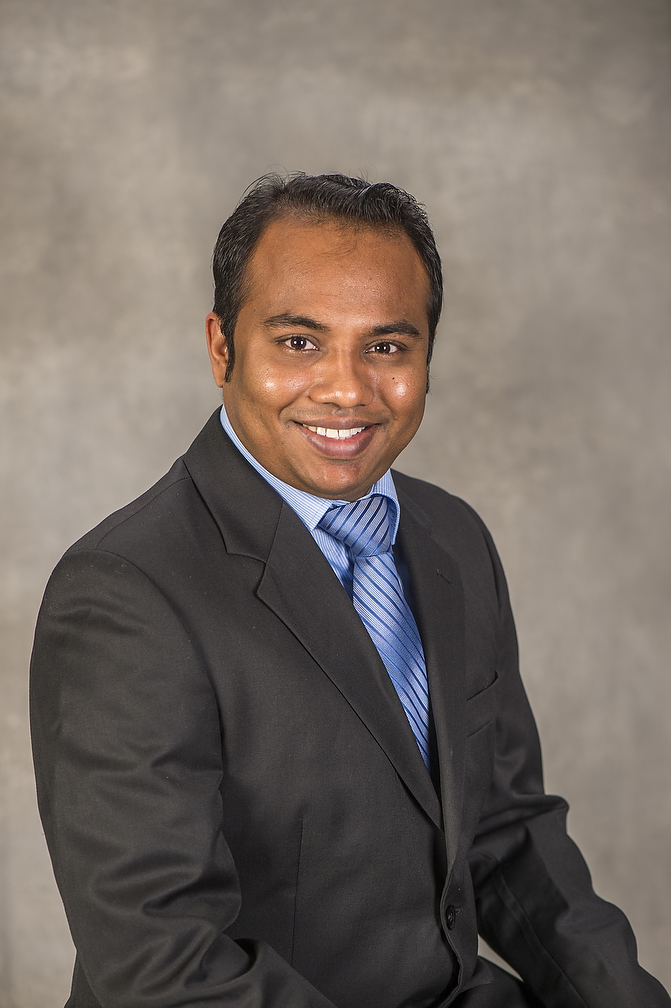}}]{Gayan Amarasuriya Aruma Baduge}   received the B.Sc. degree in  Engineering (with first class
	Hons.)  from the Department of Electronics and
	Telecommunications Engineering, University of
	Moratuwa, Moratuwa, Sri Lanka, in 2006, and the
	Ph.D. degree in Electrical Engineering from the
	Department of Electrical and Computer Engineering,
	University of Alberta, Edmonton, AB, Canada, in
	2013.  He was   a Postdoctoral Research Fellow
	with the Department of Electrical Engineering,
	Princeton University, Princeton, NJ, USA from 2014 to 2016.
	Currently, he is an assistant professor in     the Department of Electrical and Computer   Engineering  in Southern Illinois University, IL, USA.	
	His research interests include massive MIMO, millimeter-wave cellular networks, non-orthogonal multiple-access, wireless energy harvesting, cooperative MIMO relay networks, cognitive spectrum sharing and physical layer security. 
	
	He is an Associate Editor for IEEE Communications Letters. He received the best paper award in Wireless Communications Symposium at IEEE Global Communications Conference, San Diego, CA, USA, Dec. 2015. He was an Exemplary Reviewer for IEEE Communication Letters in 2011, 2012 and for IEEE Wireless Communications Letters in 2013.   He has served as a member of Technical Program Committees in several special issues for IEEE Journal on Selected Areas in Communications and at many IEEE conferences, including ICC, GLOBECOM, WCNC, PIMRC and VTC.
\end{IEEEbiography} 
\vspace{-1cm}
\begin{IEEEbiography}[{\includegraphics[width=1in,height=1.25in,clip,keepaspectratio]{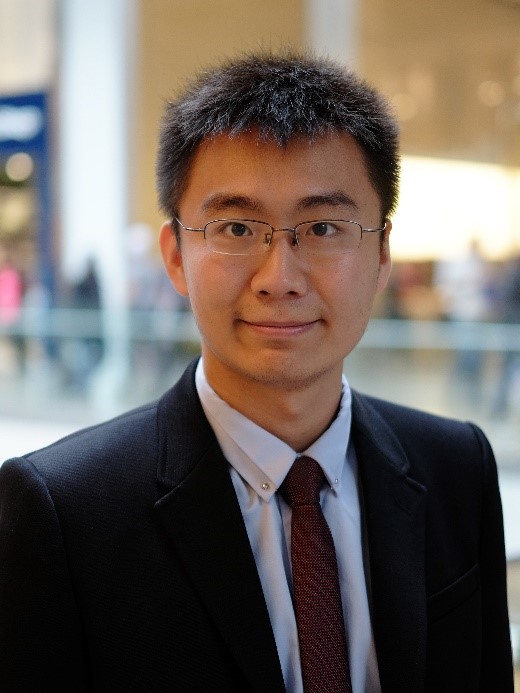}}]{Yuanwei Liu} (S'13, M'16, SM'19) received the B.S.  and M.S. degrees from the Beijing University of  Posts and Telecommunications in 2011 and 2014,  respectively, and the Ph.D. degree in electrical engineering from the Queen Mary University of London,  U.K., in 2016. He was with the Department of Informatics, King’s  College London, from 2016 to 2017, where he  was a Post-Doctoral Research Fellow. He has been a Lecturer (Assistant Professor) with the School  of Electronic Engineering and Computer Science, 
	Queen Mary University of London, since 2017. 
	
	His research interests include 5G and beyond wireless networks, the Internet of Things, machine learning, and stochastic geometry. He has served as a TPC Member for many IEEE  conferences, such as GLOBECOM and ICC. He received the Exemplary Reviewer Certificate of IEEE WIRELESS COMMUNICATIONS LETTERS in  2015, IEEE TRANSACTIONS ON COMMUNICATIONS in 2016 and 2017, and IEEE TRANSACTIONS ON WIRELESS COMMUNICATIONS in 2017 and  2018. He has served as the Publicity Co-Chair for VTC 2019-Fall. He is 
	currently an Editor on the Editorial Board of the IEEE TRANSACTIONS  ON COMMUNICATIONS, IEEE COMMUNICATIONS LETTERS, and IEEE  ACCESS. He also serves as a Guest Editor for IEEE JSTSP special issue on Signal Processing Advances for Non-Orthogonal Multiple Access in Next 
	Generation Wireless Networks.
	
\end{IEEEbiography}
\vspace{4cm}
\begin{IEEEbiography}
	[{\includegraphics[width=1in, height=1.25in, clip, keepaspectratio]{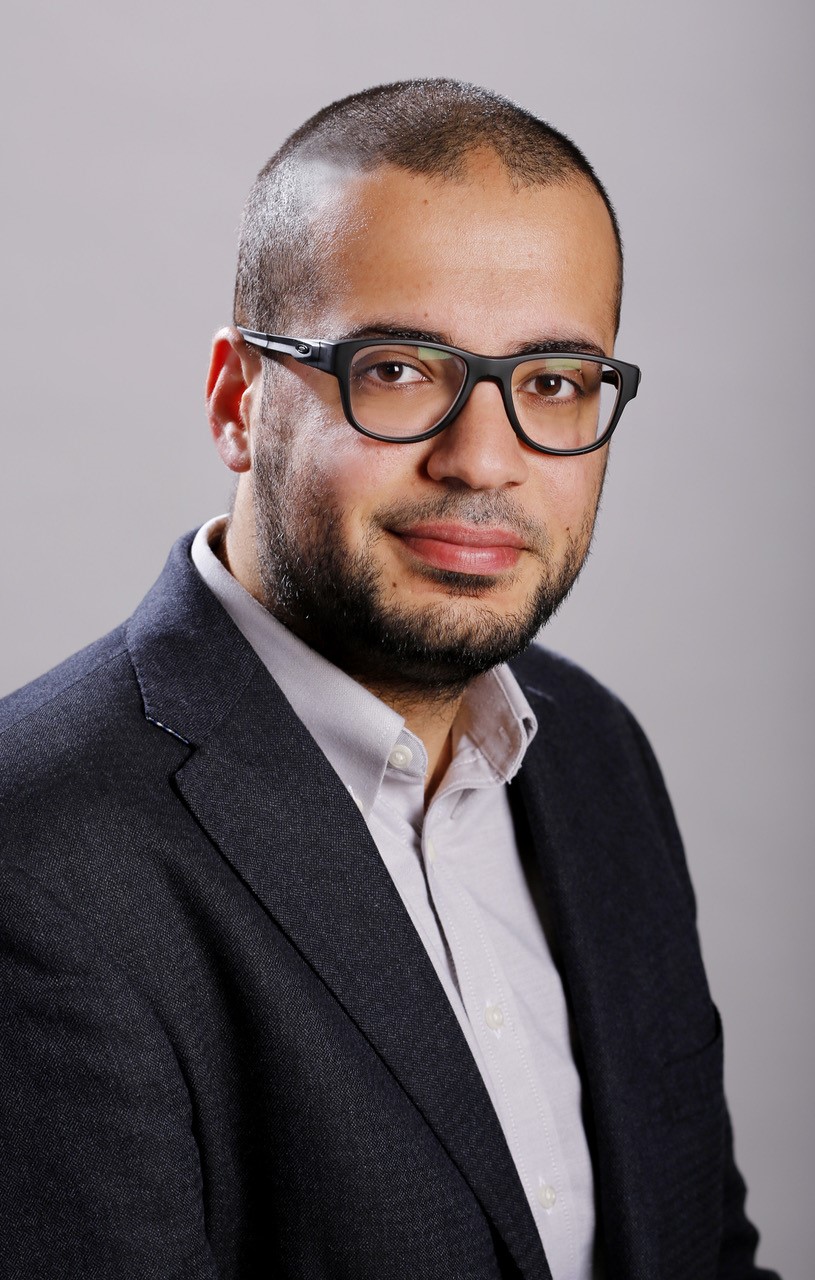}}]{Ahmed Arafa} (S'13--M'17) received the B.Sc.~degree, with highest honors, in electrical engineering from Alexandria University, Egypt, in 2010, the M.Sc.~degree in wireless technologies from the Wireless Intelligent Networks Center (WINC), Nile University, Egypt, in 2012, and the M.Sc.~and Ph.D.~degrees in electrical engineering from the University of Maryland at College Park, MD, USA, in 2016 and 2017, respectively. He has been a Postdoctoral Research Associate in the Electrical Engineering Department at Princeton University during 2017-2019. He is now an Assistant Professor in the Department of Electrical and Computer Engineering at the University of North Carolina at Charlotte.
	
	Dr.~Arafa's research interests are in communication theory, information theory, and networks, with recent focus on energy harvesting communications, age of information, physical layer security, non-orthogonal multiple access systems, and visible light communications. He was the recipient of the Distinguished Dissertation award from the Department of Electrical and Computer Engineering, University of Maryland, in 2017, for his Ph.D.~thesis work on optimal energy management policies in energy harvesting communication networks with system costs.
\end{IEEEbiography}

\vspace{-4cm}
\begin{IEEEbiography}[{\includegraphics[width=1in,height=1.25in,clip,keepaspectratio]{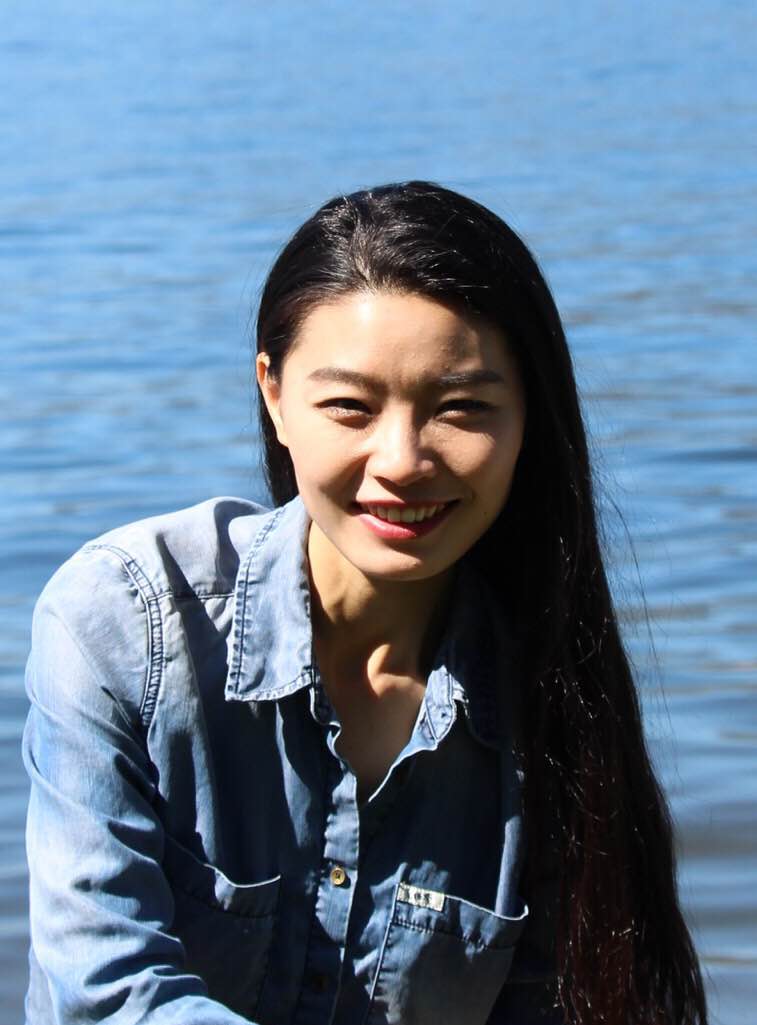}}]{Fang Fang} received the B.A.Sc. and the M.A.Sc. degrees in electronic engineering from Lanzhou
	University in 2010 and 2013, respectively, and the Ph.D. degree in electrical engineering from the University of British Columbia (UBC), Kelowna, BC, Canada, in 2018. She is currently a Post
	Doctoral Research Associate with school of electrical and electronic engineering, the University of Manchester, Manchester, U.K. Her current research interests include NOMA, machine learning and mobile edge computing. She has served as a TPC member for IEEE conferences, i.e., IEEE GLOBECOM and IEEE ICC. She received the Exemplary Reviewer Certificate of the IEEE Transactions on Communications, 2017.
	
\end{IEEEbiography}
\vspace{-4cm}

\begin{IEEEbiography}[{\includegraphics[width=1in,height=1.25in,clip,keepaspectratio]{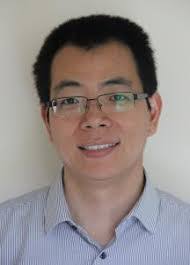}}]{Zhiguo Ding} is currently a Professor in Communications at the University of Manchester. From Sept. 2012 to Sept. 2019, he has also been an academic visitor in Princeton University. Dr. Ding’ research interests are 5G networks, signal processing and statistical signal processing. He has been serving as an Editor for IEEE TCOM, IEEE TVT, and served as an editor for IEEE WCL and IEEE CL. He received the EU Marie Curie Fellowship 2012-2014, IEEE TVT Top Editor 2017, 2018 IEEE COMSOC Heinrich Hertz Award, 2018 IEEE VTS Jack Neubauer Memorial Award, and 2018 IEEE SPS Best Signal Processing Letter Award.
\end{IEEEbiography}

\end{document}